\documentclass[aps,floats,showpacs,amstex,amssymb,bm,tightenlines,notitlepage,hidelinks,floatfix,superscriptaddress]{revtex4-2}
\def\a{\alpha}

\def\e{\epsilon}
\def\g{\gamma}
\def\w{\omega}
\def\p{\partial}

\def\M{{\cal M}}
\usepackage{amsmath}
\usepackage{xcolor}
\usepackage{enumerate}
\usepackage{comment}
\usepackage{graphicx}

\begin{document}
\title{Linear Stability of Black Holes and Naked Singularities}
\author{Gustavo Dotti}%Firstname Lastname $^{1,\ddagger}$ and Firstname Lastname $^{2,}$*}
\email{gdotti@famaf.unc.edu.ar}
\affiliation{Facultad de Matemática, Astronomía, F\'{\i}sica y Computaci\'on (FaMAF), Universidad Nacional de C\' ordoba  
and Instituto de F\'{\i}sica  Enrique Gaviola, CONICET. Ciudad Universitaria, (5000) C\'ordoba, Argentina.}

\begin{abstract} These  notes follow  from a course  delivered at the V Jos\'e Pl\'{\i}nio Baptista School of Cosmology, 
held at Guarapari (Esp\'{\i}rito Santo) Brazil, from 30 September to 
5 October 2021.  A review of the current status of the linear stability of black holes and naked singularities  is given. 
The standard modal approach,  that takes advantage of the background symmetries and analyze separately the 
harmonic components of linear 
perturbations, is briefly  introduced and used to prove that the naked singularities in the Kerr--Newman family, 
 as well as the inner black hole regions beyond Cauchy horizons, are unstable and therefore unphysical. 
The proofs require 
a treatment of  the boundary condition at the timelike boundary,  which  is given  in detail.
 The nonmodal  linear stability concept  is then introduced, and used to prove 
that the domain of outer communications of a  Schwarzschild black hole with a non-negative cosmological constant 
satisfies this stronger stability condition, which rules out transient growths of perturbations, and also to show that the 
perturbed black hole settles into a slowly rotating Kerr black hole. The encoding  of the perturbation fields in gauge invariant 
curvature scalars and the effects of the perturbation  on the geometry of the spacetime is discussed. 
\end{abstract}

\maketitle

\tableofcontents
%%%%%%%%%%%%%%%%%%%%%%%%%%%%%%%%%%%%%%%%%%
%\setcounter{section}{-1} %% Remove this when starting to work on the template.
\section{Introduction}
%we sugget to remove the contents directory before, since it is unnecessary, please confirm [DONE]

The Kerr  two-parameter family of metrics is arguably  the most important  solution of Einstein's 
equations.  These metrics represent  compact objects in an asymptotically Minkowskian, de Sitter or 
anti de Sitter background, according to the value of the cosmological constant $\Lambda$. 
 In  Boyer-Lindquist coordinates  $(t,r,\theta,\phi)$, the Kerr metric is given~by:
  \begin{multline}
ds^2=-\frac{(\Delta-a^2\sin^2\theta\Delta_{\theta})}{I^2 \Sigma}dt^2+\frac{\Sigma}{\Delta}dr^2
  +\frac{2a\sin^2\theta[\Delta-(r^2+a^2)\Delta_{\theta}]}{I^2 \Sigma}dt d\phi\\
 +\frac{\Sigma}{\Delta_{\theta}}d\theta^2+\frac{\sin^2\theta[(r^2+a^2)^2\Delta_{\theta}-a^2\sin^2\theta\Delta]}{I^2 \Sigma}d\phi^2,
  \label{kerr}
\end{multline}
where $(\theta,\phi)$ are the standard angular coordinates of the unit sphere, $\Lambda$ is the cosmological constant and 
we introduced the notation 
\begin{equation}\label{kerr2}
\begin{split}
 \Delta &= r^2-2Mr+a^2-\tfrac{1}{3}\Lambda r^2(r^2+a^2), \\
  I&= 1 + \tfrac{1}{3} \Lambda a^2, \\
 \Delta_{\theta} &= 1+\tfrac{1}{3}\Lambda a^2\cos^2\theta, \\ 
 \Sigma &= r^2+a^2\cos^2\theta. 
 \end{split}
\end{equation}

For large $r$, Equation (\ref{kerr})  %please consider this change  [AGREED]
approaches the metric of Minkowski spacetime  if $\Lambda=0$, de Sitter if $\Lambda$ is positive and 
  anti-de Sitter if negative. 
The parameters $(M,a)$ appear as constants of integration and thus  can  take any value. 
They can be  seen to correspond, respectively, to the mass and the angular momentum per 
unit mass of these objects.  If $a<0$, we may change coordinates $\phi \to -\phi$ and get the same form 
(\ref{kerr}) with $a \to -a$. So we may assume without loss of generality that $a \geq 0$. 
All these spacetimes contain a curvature singularity. This is milder  in the rotating, $a > 0$ case: it lies at 
$r=0$, $\theta=\pi/2$ and it is 
 possible to  go across it to negative $r$ values \cite{onil}.
 %wrong ref order, please revise all the ref citation and keep them in a numericla and sequential order throughout the whole manuscript

In the non-rotating $a=0$ case, Kerr's metric reduces to Schwarzschild's:
\begin{equation} \label{schwa}
ds^2 =-\left( 1 - \frac{2M}{r} \right) dt^2 + \frac{dr^2}{\left( 1 - \frac{2M}{r} \right)} + r^2 \left( d\theta^2 + \sin^2 \theta \ d\phi^2 \right),
\end{equation}
 for which   the singularity is at $r=0$ and signals a boundary of the spacetime.

For certain combinations of $M,a$ and $\Lambda$, one or two horizons keep the singularity causally isolated from the 
external, asymptotically simple region where the metric approaches that of 
Minkowski or (anti) de Sitter space, called  the Domain of Outer Communications (DOC): 
these are  the black hole (BH)  solutions. For other combinations the singularity is visible from the DOC: these are 
called \textit{naked singularities} (NSs).

Consider, for example, the asymptotically Minkowskian case $\Lambda=0$. 
The horizons correspond to  the hypersurfaces given by the roots of $\Delta$ in (\ref{kerr2}) \cite{onil}:
\begin{equation}
r_\pm = M \pm \sqrt{M^2-a^2}.
\end{equation}

We will focus 
on the following cases:

\begin{itemize}
\item Rotating Kerr metric ($a>0$): sub extreme BH ($M>a$, two horizons, at $r=r_\pm$); extreme BH  ($M=a$, a single  horizon at $r=r_+=r_-=M$); 
super-extreme NS ($0<M<a$, no horizon).
\item Non-rotating Schwarzschild metric ($a=0$): Schwarzschild BH ($M>0$, one horizon at $r=2M$), Schwarzschild NS ($M<0$, no horizon).

\end{itemize}

When analyzing geodesics we find that the Schwarzschild NS effectively behaves as a central object with a negative mass: it is 
a totally unphysical solution of Einstein's vacuum equations. The rotating NS $a>M>0$ 
has a  causal behavior that is completely pathological: through any point there 
passes a closed timelike curve \cite{onil}. The sub-extreme Kerr BH, on the other hand, is among the physically most relevant 
solutions of Einstein's equation. This is so in view of the uniqueness theorems \cite{Robinson:1975bv,heusler}, that state that the $\Lambda=0$ 
Kerr's BH is \textit{{the only}} %if the italic necessary, the same with all
asymptotically Minkoswkian rotating BH solution of Einstein's equations. 
Black holes are the most extraordinary prediction of Einstein's gravity, and if  sixty years ago 
we were beginning to \textit{mathematically} understand them, in 2015, with the first detection
of gravitational waves from BH collisions, we entered an era of 
\textit{direct observation} of these objects. We are now potentially  able to test whether or not the $r>r_+$ piece of the 
$M>a>0$ Kerr metric (\ref{kerr}) correctly models the spacetime outside a rotating BH.

One may wonder if the existence of  unphysical solutions such as the Kerr NS is a flaw of General Relativity. 
The answer is in the negative. A close analogy can be found in Classical Mechanics: Kerr's solution 
(\ref{kerr}) is \textit{stationary}, roughly meaning that the space geometry is the same at all times (technically: 
the metric is invariant under the flow of the vector field $\p/\p_t$), so it is analogous to time-independent 
solutions in Classical Mechanics, which can be characterized as the system sitting at a critical point of the potential 
energy function. 
Critical points can be local minima or not, and it is only in the first case that the time-independent solution is feasible. 
A regular cone ``resting'' vertical on its tip is a solution of Newton's equations, but we never 
encounter these configurations. The reason is that these are \textit{unstable} time-independent solutions, 
they require fine tuning of the initial condition and under the slightest 
perturbations the system evolves into a very different configuration. 
In the following sections we will show that this is exactly the case of the NSs in the Kerr $(a,M)$ family of solutions.

For sub-extreme rotating BHs, the inner horizon $r_-$ is also a \textit{Cauchy horizon}: a null hypersurface beyond which  
the uniqueness of the evolution is lost. Although there is a unique \textit{analytic} continuation of the metric beyond the Cauchy horizon,  
there are \textit{infinitely}  
many possible \textit{smooth} continuations satisfying Einstein's vacuum equation.
This is an annoying feature, contrary to the central idea in Classical Physics that we can  predict uniquely the evolution of a system 
from initial data. These inner BH regions also exhibit severe causal pathologies, such as closed timelike curves. 
We will prove below that  these spacetime regions are unstable under perturbations, 
that is, unphysical extensions of the BH metric.

Having found an explanation for the unphysical solutions, we would like to make sure that the  
solutions that  model BH exteriors, that is the $r>r_+$ piece of the $M \geq a \geq 0$ Kerr and Schwarzschild metrics,  
{are stable under perturbations}. 
Otherwise they  would be  irrelevant solutions  
of General Relativity.  
 We have entered an era where the strong gravity regime of BH can be tested by means of gravitational waves,  
  multi-messanger Astronomy and direct observations of BH horizons. General Relativity is the available theory of gravity, 
  and thus  Kerr solution is used to model the BH exterior region. 
  If a mismatch is observed between theory and experiment, in view of the uniqueness theorems, 
  corrections to General Relativity will have to be considered.

A complete proof of the stability of Kerr BH using the full, non-linear Einstein's equation is and will be lacking for a time. 
We should recall at this point that such a proof only exists for: (i) de Sitter spacetime (proved by H. Friedrich in 1986 \cite{f,f2}), (ii) 
Minkowski spacetime (proved by Christdoulou and Klainerman in 1993 \cite{ck}), (iii) Schwarzschild de Sitter BH, 
(proved by Hintz and Vasy in 2016 \cite{hv}).  A  preprint  is available since a few months ago  with 
a proof of non-linear stability   of the  $\Lambda=0$ Schwarzschild  BH (see  \cite{Daf2}), the proof takes five hundred pages. 
For the  Kerr BH,  at the moment, we must content ourselves with a proof of \textit{linear stability}.

Linear perturbations around a metric can be organized in \textit{modes} by taking advantage of the symmetries of the background: perturbations 
are classified according to the way they transform under the action of the isometry group 
 of the unperturbed metric. Single modes are found to either oscillate  or 
 grow/damp exponentially in time. If a single unstable---that is, exponentially growing---mode is found, the background metric is certainly unstable. 
 The existence of such modes is what allows us to rule out NS and pathological BH interiors, as proved below. 
The absence of unstable modes on  BH exteriors 
 is  a solid signal of stability: we call this \textit{modal linear stability}. There is still the possibility that
  oscillating modes add up to a locally growing  perturbation. Ruling out this possibility is what we call 
  \textit{nonmodal linear~stability}.

  This review is organized as follows: in Section \ref{Slm} we briefly introduce the ideas of mode decomposition 
  of linear perturbations based on background symmetries; in \mbox{Section \ref{Sus}} we review the proofs of instability of the Schwarzschild and Kerr NS instabilities,  
  and also of the instability of the region beyond the Cauchy horizon of Kerr BHs; in Section \ref{nms} we review the proof of 
  nonmodal stability of the Schwarzschild (Schwarzschild de Sitter) BH. In doing this we show that there are 
 gauge invariant  scalar fields, constructed out of the perturbed curvature scalars (CSs), that is,  
 contractions of the Weyl tensor and its first covariant derivative, 
  that contain the exact same information as the gauge class of the metric perturbation, and use them to 
  show that, for the most general perturbation, no transient growths  are possible, and that a perturbed 
 Schwarzschild  BH always decays into a slowly rotating Kerr BH. The discussion is centered on 
 the effect of perturbations on the background geometry.

\section{Mode Decomposition of Linear Perturbations} \label{Slm}

Linear perturbation theory assumes that,  on a fixed four dimensional manifold $\M$---the spacetime---there is a 
 mono-parametric family of solutions $g_{\alpha \beta}(\e)$ of the vacuum Einstein field equation
 \begin{equation} \label{mpfs}
R_{\alpha \beta}[g(\e)] - \Lambda g_{\alpha \beta}(\e)=0
\end{equation}
around the ``unperturbed background" $g_{\alpha \beta}=g_{\alpha \beta}(0)$. 
A Taylor expansion of (\ref{mpfs}) at $\e=0$ keeping up to first order terms implies that  the 
\textit{ metric perturbation} 
\begin{equation} \label{mp}
h_{\alpha \beta}= \left. \frac{d}{d \epsilon} g_{\a \beta }(\e)  \right|_{\e=0},
\end{equation}
satisfies  the linearized Einstein Equation (LEE) 
\begin{equation} \label{lee}%we suggest to  change it into number  [AGREED]
{\cal E}[ h_{\alpha \beta}] := - \tfrac{1}{2} \nabla^{\gamma } \nabla_{\gamma } h_{\alpha \beta }  - \tfrac{1}{2} \nabla_{\alpha } \nabla_{\beta }
 (g^{\gamma \delta } h_{\gamma \delta }) +
  \nabla^{\gamma } \nabla_{(\alpha } h_{\beta) \gamma } - \Lambda \;  h_{\alpha \beta } =0. 
\end{equation}

Trivial solutions of this equation  are 
\begin{equation} \label{puregauge}
 h_{\alpha \beta} =  \pounds_{\xi}   g_{\alpha \beta} = \nabla_{\alpha} \xi_{\beta} +
\nabla_{\beta} \xi_{\alpha},
\end{equation}
where $\xi^{\a}$ is an arbitrary  vector field: these amount to 
the first order change of the metric under the flow  generated by  the vector field $\xi^{\a}$. 
 Any two
 solutions $h_{\alpha \beta}$ and $h'_{\alpha \beta}$ of the LEE such that 
\begin{equation} \label{gt2}
h_{\alpha \beta}' = h_{\alpha \beta} +  \pounds_{\xi}   g_{\alpha \beta},
\end{equation}
 are related by this diffeomorphism, and then physically equivalent. This is   the {\em gauge freedom} of linearized gravity, it can be used 
 to set the metric perturbation into a particular form, in the same way gauge freedom is used in electromagnetism to impose  conditions on the 4-vector potential.

\subsection{Schwarzschild Background}
Complicated equations admit 
 relatively simple solutions under highly symmetric ansatz. This is how 
Schwarzschild solution was obtained a few months after the publications of Einstein's 
equations: a non-linear second order partial differential equation  involving 10 functions was reduced, 
under the static, spherically symmetric metric  ansatz, 
 to a
system of 2 ordinary differential equation . 
When linearizing the equations around such a simple solution, the 
background symmetries ``organize'' the set of solutions of the linearized equations. 
If $g_{\alpha \beta}$ is a solution of Einstein's equations and $h_{\alpha \beta}$ a solution 
of the Linearized Einstein's Equations (\ref{lee}) %%we suggest to  change it into number
around $g_{\alpha \beta}$, then, under the action of 
an isometry, $g_{\alpha \beta}+\e h_{\alpha \beta} \to g_{\alpha \beta}+\e \tilde h_{\alpha \beta}$, with $\tilde h_{\alpha \beta}$ a different solution of the LEE.  
The  solutions of the LEE live in representation spaces  of the isometry group and can be classified into invariant subspaces 
according to their Casimir values. For example, 
the isometry group of Schwarzschild solution 
is 4 dimensional: $P \times SO(3) \times \mathbb{R}_t$: rotations, time translations, and 
the discrete parity transformation $P(t,r,\theta,\phi)=(t,r,\pi-\theta,\phi+\pi)$. Infinitesimal  rotations 
give rise to the Killing vector fields
\begin{equation}
\begin{split}
J_1 &= -\sin \phi \p_\theta- \cot \theta \cos \phi \p_\phi\\
J_2 &= \cos \phi \p_\theta -\cot \theta \sin \phi \p_\phi\\
J_3&= \p_\phi.
\end{split}
\end{equation}

Perturbations can be classified according to their behavior under the square angular momentum operator
\begin{equation}
\mathbf{J}^2 = \sum_{m=1}^3 (\pounds_{J_{(m)}})^2
\end{equation}
and the pull-back $P_*$ under parity. A general perturbation 
can be written as 
\begin{equation} \label{gp}
h_{\alpha \beta} = \underbrace{ \sum_{\ell=2}^\infty\sum_{m=-\ell}^\ell h_{\alpha \beta}^{(\ell,m,-)}}_{\text{odd modes}}
+ \underbrace{\sum_{\ell=2}^\infty\sum_{m=-\ell}^\ell h_{\alpha \beta}^{(\ell,m,+)}}_{\text{even modes}}
\end{equation} where 
 the $(\ell,m,\pm)$ modes satisfy 
\begin{align}
\mathbf{J}^2  h_{\alpha \beta}^{(\ell,m,\pm)} &= - \ell (\ell+1) h_{\alpha \beta}^{(\ell,m,\pm)}, \\
 P_* h_{\alpha \beta}^{(\ell,m,\pm)} &= \pm (-1)^{\ell} 
h_{\alpha \beta}^{(\ell,m,\pm)}. \label{ps}
\end{align}

Modes with $\ell=0,1$ can be omitted in (\ref{gp}) in an stability study because they are non-dynamical, and therefore irrelevant to 
the stability problem, they can be shown to be either pure gauge or 
infinitesimal displacement within the Kerr family: changes of the background mass, or addition of angular momentum 
(see, e.g., \cite{dottiCQG})).

The Ricci tensor admits a decomposition like (\ref{gp}), and it can be proved that the component $\delta R_{\alpha \beta}^{(\ell,m,\pm)}$ 
of $\delta R_{\alpha \beta} \equiv  \left. \frac{d}{d \epsilon} R_{\a \beta }(\e)  \right|_{\e=0}$ 
depends only of $h_{\alpha \beta}^{(\ell,m,\pm)}$ (same $(\ell,m,\pm)$). In other words, different modes decouple, so we can 
solve the LEE for single modes.

The explicit form of the odd modes in Regge-Wheeler gauge is
\begin{equation}\label{odd1}
h_{\alpha \beta}^{(\ell,m,-)} = \left( \begin{array}{cc} 0 & A^{(\ell,m)} \\ A^{(\ell,m)} & 0 \end{array} \right) \\
\end{equation}
where 
\begin{equation}\label{odd2}
 A^{(\ell,m)} = \left( \begin{array}{cc} H_t(t,r) \frac{1}{\sin \theta} \p_\phi S^{(\ell,m)}(\theta,\phi) & - H_t(t,r) \sin \theta \p_\theta S^{(\ell,m)}(\theta,\phi) \\
H_r(t,r) \frac{1}{\sin \theta} \p_\phi S^{(\ell,m)}(\theta,\phi)  & - H_r(t,r) \sin \theta \p_\theta S^{(\ell,m)}(\theta,\phi) \end{array} \right)
\end{equation}
and $S^{(\ell,m)}(\theta,\phi) $ are orthonormal real spherical harmonics:
\begin{equation} \label{sh}
\frac{1}{\sin \theta} \p_\theta ( \sin \theta \; \p_\theta S^{(\ell,m)} ) + \frac{1}{\sin^2 \theta} \p_\phi^2 S^{(\ell,m)} = -\ell (\ell+1) S^{(\ell,m)}.
\end{equation}

For the even modes

\begin{equation}\label{e1}
h_{\alpha \beta}^{(\ell,m,+)} = \left( \begin{array}{cc} B^{(\ell,m)} & 0  \\0 &  C^{(\ell,m)}   \end{array} \right)
\end{equation}
where, in terms of  $f(r)=1-2M/r$, 
\begin{equation}\label{e2}
 B^{(\ell,m)} = \left( \begin{array}{cc}f(r) H_0(t,r) S^{(\ell,m)}(\theta,\phi) & H_1(t,r) S^{(\ell,m)}(\theta,\phi) \\
H_1(t,r)  S^{(\ell,m)}(\theta,\phi) &\frac{1}{f(r)} H_2(t,r) S^{(\ell,m)}(\theta,\phi)  \end{array} \right)
\end{equation}
and 
\begin{equation}\label{e3}
 C^{(\ell,m)} = \left( \begin{array}{cc}r^2 K(t,r) S^{(\ell,m)}(\theta,\phi)  & 0 \\ 0 & r^2 \sin^2\theta K(t,r) S^{(\ell,m)}(\theta,\phi)\end{array} \right).
\end{equation}

Introducing the Regge-Wheeler fields $\phi^{(\ell,m,-)}$ defined by (refer to (\ref{odd1}) and (\ref{odd2}))
\begin{equation}
\begin{split}
H_t(t,r)&=-f(r)\left( \phi^{(\ell,m,-)} + r \p_r \phi^{(\ell,m,-)}\right), \\
H_r(t,r)&=-\frac{r}{f(r)} \p_t  \phi^{(\ell,m,-)},
\end{split}
\end{equation}
and the Zerilli field  $\phi^{(\ell,m,+)}$, defined by (see (\ref{e1})--(\ref{e3}))
\begin{equation}
\label{RW2}
\begin{split}
K & = q(r) \phi^{(\ell,m,+)}+\left(1-\frac{2M}{r}\right)\frac{\partial \phi^{(\ell,m,+)}}{\partial r}, \\
H_1 & =  h(r)\frac{\partial \phi^{(\ell,m,+)}}{\partial t}+r\frac{\partial^2 \phi^{(\ell,m,+)}}{\partial t \partial r},
 \\
H_2 & =  \frac{\partial }{\partial r}\left[\left(1-\frac{2M}{r}\right)\left(h(r) \phi^{(\ell,m,+)},
+r\frac{\partial \phi^{(\ell,m,+)}}{ \partial r} \right)\right] -K , 
\end{split}
\end{equation}
where
\begin{equation}
\label{RW2a}
\begin{split}
\lambda &= \frac{(\ell-1)(\ell+2)}{2}\\
q(r) & =   \frac{\lambda(\lambda+1)r^2+3 \lambda M r + 6 M^2}{r^2(\lambda r + 3 M)}, \\
h(r) & =   \frac{\lambda r^2 - 3 \lambda r M -3 M^2}{(r-2M)(\lambda r + 3M)},
\end{split}
\end{equation}
the full set of linearized Einstein's equations  reduce to: (i)  $H_0=H_2$ in (\ref{e2}),  
(ii)  the spherical harmonic Equation (\ref{sh})  
and (iii) the  Regge-Wheeler (odd modes) and Zerilli (even modes) 1 + 1 wave equations below:
\begin{equation}\label{1+1}
[\p_t^2 \underbrace{- \p_x^2 + V^{(\ell,m,\pm)}}_{\mathcal{H}^{(\ell,m,\pm)}} ] \phi^{(\ell,m,\pm)} =0
\end{equation}
where $x$ is a ``tortoise'' radial coordinate, defined by $dx/dr=1/f(r)$, for which 
 we choose the integration constant such that  
\begin{equation}
\begin{split}\label{x}
x& =r+2M \ln\left (\frac{r-2M}{2M}\right), \;\;\; \text{$r>2M, M>0$ (BH)} \\
x&= r+2M \ln\left (\frac{r-2M}{-2M}\right), \;\;\; \text{$r>0,M<0$ (NS)}
\end{split}
\end{equation}

Note that  the 1+1 wave Equation (\ref{1+1}) is defined on a two dimensional Minkowski space in the BH case,  
and a half of that space  for the NS, since 
\begin{equation}\label{x2}
\begin{split}
-\infty < x < \infty , \;\;\; \text{$M>0$ (BH)} \\
0 < x <\infty, \;\;\; \text{$M<0$ (NS).} 
\end{split}
\end{equation}

This peculiar domain for the NS Equation (\ref{1+1}) is connected with the fact that the singularity is \textit{timelike} (unlike in the BH case): 
 we have  to analyze which boundary conditions should we impose to the perturbation at the singularity.\\
The even and odd potentials in (\ref{1+1}) are 

\begin{equation}
\begin{split}
\label{poten1}
V^{(\ell,m,+)} &= 2 f(r) {\frac{ \lambda^2 r^2
\left[(\lambda+1) r + 3 M \right] + 9 M^2 (\lambda r +M)}{ r^3
(\lambda r + 3 M)^2} }\\
V^{(\ell,m,-)} &=f(r) \left(\frac{\ell (\ell+1)}{r^2} - \frac{6M}{r^3} \right)
\end{split}
\end{equation}

where, in each case,  we should replace $r \to r(x)$ given by the inverse of (\ref{x}).\\

The \textit{modal linear stability} of the Schwarzschild BH now follows in a few steps:

\begin{itemize}

\item Equation (\ref{1+1}) admit separable solutions of the form 
\begin{equation} \label{sep}
\phi^{(\ell,m,\pm)}=e^{\pm i \omega t} \psi^{(\ell,m,\pm)}(x)
\end{equation}

with $\psi^{(\ell,m,\pm)}(x)$ an eigenfunction of the  operators defined in (\ref{1+1}), which 
have the form of a quantum Hamiltonian:

\begin{equation}\label{qh}
\mathcal{H}^{(\ell,m,\pm)} \psi^{(\ell,m,\pm)} = \w^2 \psi^{(\ell,m,\pm)}, \;\; -\infty < x < \infty
\end{equation}

\item For BH the solutions of (\ref{1+1})   should be square integrable in $x$ for properly decaying metric perturbations, 
the stability problem then reduces to determining if \textit{any} of the Hamiltonians $\mathcal{H}^{(\ell,m,\pm)}$ admits a 
negative energy, bounded eigenfunction, as in this case $\w=\pm i k$, $k \in \mathbb{R}$ would cause exponentially growing 
solutions of the form (\ref{sep})
\item The potentials $V^{(\ell,m,\pm)}$ are indeed positive definite and decaying for $|x| \to \infty$, the spectra of 
the $\mathcal{H}^{(\ell,m,\pm)}$ is positive definite. 
Exponentially growing modes are, therefore, ruled out.
\end{itemize}

\subsection{Kerr Background}

 Kerr's solution has a much smaller isometry group than Schwarzschild's: $U(1)\times \mathbb{R}_t$ 
($e^{i \alpha} \in U(1)$ operates by sending $(t,r,\theta,\phi)  \to (t,r,\theta,\phi+\alpha)$, $T \in \mathbb{R}_t$ by sending 
$(t,r,\theta,\phi)  \to (t+T,r,\theta,\phi+\alpha)$t, as in Schwarzschild). Its 
perturbative analysis is far more complicated than that of the Schwarzschild metric because the equations involving the metric perturbation are, 
as a consequence of the reduced symmetries, non separable. 
 The first, and still the most successful approach to linear perturbations 
of the Kerr metric was developed by Teukolsky in \cite{teukolsky,teukolsky2} and is based on  the Newman-Penrose null tetrad approach, 
 as we know briefly explain.

 The Weyl tensor of a generic metric is of type-I in the Petrov classification. This means that the eigenvalue problem 
 \begin{equation} \label{ev-i}
\tfrac{1}{2} C^{\a \beta }{}_{\g \delta } X^{\g \delta} = \lambda X^{\alpha \beta}, \;\; X^{\a \beta}=X^{[\a \beta]}
\end{equation}
admits three {\em different} solutions, with $\lambda_1+\lambda_2+\lambda_3=0$ or, equivalently, that the equation 
\begin{equation}\label{pnd-I}
 k_{[\mu} C_{\a]\beta \g[\delta}k_{\nu]}k^{\beta}k^{\g}=0, \;\;\; k^\a k_\a=0,
\end{equation}
admits four solutions spanning four different null {\em lines} (called principal null {\em directions} or PNDs).
Type D spacetimes, instead, are characterized by the fact that the eigenvalue Equation (\ref{ev-i}) admits three linearly independent
 solutions with $\lambda_1=\lambda_2=-\tfrac{1}{2} \lambda_3$, 
a condition that turns out to be equivalent to the existence of two  double PNDs---that is, two non-proportional null vectors 
satisfying 
an equation stronger than (\ref{pnd-I}):
\begin{equation}\label{pnd-d}
 C_{\a \beta \g[\delta}k_{\mu]}k^{\beta}k^{\g}=0, \;\;\; k^\a k_\a=0.
\end{equation}

It should be stressed that  Equations (\ref{pnd-I}) and (\ref{pnd-d}), being homogeneous, do not define (null) 
tangent vectors at a point $p$ of the spacetime  but instead one dimensional subspaces since, given a solution $k^\a$, 
$z k^\a$ is also a  solution for any $z$. \\

 All metrics in the Kerr family (and more generally, in the Kerr--Newman family of charged and/or rotating BHs) 
 have  type D metrics. We may choose two null vectors $l^\a$ and $n^\a$ generating the PNDs to be future pointing   
and normalized to $l^\a n_\a=-1$. These vectors span a two dimensional  $(-,+)$ subspace of the tangent space at each point 
whose orthogonal complement is a two dimensional $(+,+)$ subspace spanned, say, by two orthonormal vectors $x^{\a }$ and $y^{\a }$.
In the  Newman-Penrose approach these are  replaced by a complex null vector $m^\a=x^\a+i y^\a$. The ambiguity in choosing $x^{\a }$ and $y^{\a }$ 
 is given by the possibility of rotating them in the plane they generate: $m^{\a } \to e^{i\beta} m^{\a }$,  
 the ambiguity in the choice of $l^{\a }$ and $n^{\a }$ reduces to  a scaling by a positive $A$: $l^{\a } \to Al^{\a }$, $n^{\a } \to A^{-1}n^{\a }$. 
 These are the only Lorentz transformations of the \textit{null tetrad} 
 $\{ l^{\a }, n^{\a }, m^{\a }, \bar m^{\a }\}$  that 
 preserve the future PNDs and the inner products $l^\a n_\a=-1, m^\a \bar m_\a=1$ (note that all other  inner products vanish).  
Tensor components in Newman-Penrose tetrads carry {\em spin weight}, a quantity that measures how many $m^a$'s and 
$\bar m^a$'s are involved in the component. For example, under $m^\a \to e^{i \beta} m^\a$, the five complex \textit{Weyl scalars} 
\begin{equation} \label{np1}
\begin{split}
\Psi_0 &= C_{\alpha \beta \gamma \delta } l^\alpha m^\beta l^\gamma m^\delta, \\
\Psi_1 &=C_{\alpha \beta \gamma \delta  }
 l^\alpha n^\beta l^\gamma m^\delta, \\
  \Psi_2 &=C_{\alpha \beta \gamma \delta} l^\alpha m^\beta \bar m ^\gamma n^\delta, \\
\Psi_3 &=C_{\alpha \beta \gamma \delta} l^\alpha n^\beta \bar m^\gamma n^\delta, \\
 \Psi_4 &=C_{\alpha \beta \gamma \delta} 
n^\alpha \bar m^\beta n^\gamma \bar m^\delta,
\end{split}
\end{equation}
transform as $\Psi_k \to e^{i (2-k) \beta} \Psi_k$, and so 
 are said to have {\em spin weight } $s=2-k$. Tetrad-dependent scalar fields such as these 
 are called \textit{spin-weighted} scalars. The five spin weighted scalars $\Psi_k$ above 
 contain all the information on the ten independent real components of the Weyl tensor. 
As another example, the six real components  of  a test electromagnetic field on a type D background can be encoded in three complex fields:

\begin{equation}\label{np2}
\phi_0 = F_{\alpha \beta} l^\alpha m^\beta, \;\; \phi_1 = \tfrac{1}{2} F_{\alpha \beta} (l^\alpha n^\beta+ \bar m^\alpha m^\beta ), 
\;\; \phi_1 = F_{\alpha \beta} \bar m^\alpha n^\beta.
\end{equation}

These carry spin weight $1,0$ and $-1$, respectively.

The systematic analysis of the linearized perturbations of Kerr space-time was greatly facilitated by Teukolsky's discovery 
\cite{teukolsky} that the LEE imply  second order {\em separable} partial differential evolution equations for some  (weighted) first order variations  
of the extreme spin weight components  \cite{teukolsky2}:
\begin{equation} \label{teukg}
\Phi_{s=2} = \delta \Psi_{0}, \;\;\;  \Phi_{s=-2}= \Psi_2 ^{-4/3} \delta \Psi_4
\end{equation}

Here $ \delta \Psi_{k}$ is the first order variation of a  Weyl scalar and  $\Psi_{k}$ its background value. 
The equation for  (weighted) extreme $s=\pm 1$ components $\Phi_{s=0}=\phi_0$ and 
$\Phi_{s=-1}=\Psi_2^{-2/3} \phi_2$ of a   test electromagnetic field (\ref{np2}) assume the same universal form as that for the 
gravitational fields (\ref{teukg}), and 
so does  the massless scalar field (for which $s=0$). After separation of~variables,

\begin{equation}\label{teukofield}
\Phi_s=
R_{\w,m,s}(r) S^{(\ell,m)}_{\w,s}(\theta) \exp(im\phi) \exp(-i\omega t),
\end{equation}
this universal equation can be given as 
  a coupled system for $S^{(\ell,m)}_{\w,s}$ and
$R_{\w,m,s}$ which, dropping indices, reads
\begin{equation} \label{ta}
{1\over \sin\theta} {d \over d\theta}\left(\sin\theta {d S\over
d\theta}\right)+\left(a^2\omega^2\cos^2\theta-2 a \omega s \cos\theta -{(m + s \cos\theta)^2\over
\sin^2\theta}  +\textcolor{black}{E} -s^2\right)S =0, 
\end{equation}

\begin{multline}
\label{tr}
\Delta {d^2 R \over dr^2} +(s+1) {d\Delta\over dr} \;{dR\over dr} +\left\{
{K^2-2is(r-M)K\over \Delta}\right. \\  \left. +4ir\omega s -[\textcolor{black}{E}-2am\omega+a^2
\omega^2-s(s+1)]\right\}R =0.
\end{multline}

Here, $s$ is the spin weight ($s=\pm 2$ in the case of gravitational perturbation, $s=\pm 1$ for test electromagnetic fields, $s=0$ for scalar massless fields), 
 $\Delta$ is given in (\ref{kerr2}) and $K=(r^2+a^2)\omega -a
m$.  
The functions $S^{(\ell,m)}_{\w,s}(\theta) e^{im\phi}$ have to be smooth on the  sphere. This gives 
 a discrete spectrum of the eigenvalues $E_\ell$, $\ell=0,1,2, ...$ in Equation (\ref{ta}) --as 
is the case for the spherical harmonic Equation (\ref{sh}){--}. 
In the spherical harmonic case ($\w=0,s=0$), the spectrum is  
 $E_\ell=\ell (\ell+1)$. In the general case of (\ref{ta}), the regular solutions $S^{(\ell,m)}_{\w,s}(\theta) e^{im\phi}$ 
 are called \textit{spin weighted spheroidal harmonics}, 
and the eigenvalues $E_\ell$ depend on the  $a\w$ and $s$, as is evident from Equation (\ref{ta}).

 Teukolsky's equations have been applied to a wide range of problems involving 
  gravitational, scalar field and Maxwell field  perturbations, as well as black hole collisions in the so called ``close limit''. 
 The $s=\pm2 $ equations are key to establishing the modal linear stability of the DOC of the Kerr  black hole under
  gravitational perturbations. This was done through a series of papers  starting from \cite{teukolsky,stable} (for the current status see \cite{lars}).

\section{Instability of Naked Singularities and Black Hole Inner Regions} \label{Sus}

In this section, we use the modal approach to the LEE developed in the previous section 
to prove the instability of NSs and of the regions beyond the Cauchy horizon in BHs. 
In Section \ref{Sisns}, we review the proof of instability of the Schwarzschild NS given in \cite{Dotti:2008ta,Gleiser:2006yz} 
  (see also \cite{ghi}), and briefly mention 
the  proof of instability of the Reissner-Norsdtröm NS. In Section \ref{Siks}, we 
review  the proof of the instability of the super-extreme Kerr NS, which is given in 
\cite{dgr,Dotti:2011eq}. Finally, in Section \ref{Sibch} we review the proof of 
instability in the inner regions beyond the Cauchy horizon of the Kerr BH \cite{doglepu, Dotti:2010uc, dgr}. 

\subsection{Instability of the Schwarzschild Naked Singularity} \label{Sisns}

When analyzing the LEE for the $M<0$ Schwarzschild NS 
we face the problem of solving a 1 + 1 wave equation on the  \textit{half} of Minkowski space $(t,x)$ corresponding to $x>0$ 
(see Equations (\ref{1+1})--(\ref{x2})). 
This is a consequence of the fact that the Schwarzschild solution with $M<0$ fails to be globally hyperbolic. 
The coordinates $-\infty<t<\infty$ and $r>0$ (together with the angular coordinates 
$(\theta,\phi)$ on the sphere) are \textit{global}, the spacetime admits no extension. 
The singularity at $r=0$ is \textit{timelike} in this case (as opposed to 
the BH case, for which is spacelike), as can be easily seen by noting that the induced metric on a hypersurface $r=r_o \simeq 0$ 
is 
\begin{equation}
ds^2 \simeq \frac{2 M}{r_o} dt^2 + r_o^2 ( d\theta^2+ \sin^2 \theta \, d\phi^2),
\end{equation}
which, given that $M<0$, has signature $(-,+,+)$. Since Schwarzschild NS fails to be   globally hyperbolic, it has no Cauchy surface 
(a surface that \textit{every} causal curves intersects exactly once). The closest to  suitable  spacelike surfaces for 
an initial value formulation 
are those defined by $t=constant$. For a field satisfying a wave equation we can give initial values of the field and its time derivative 
at such a surface. However, the evolution will not be entirely defined unless we  add suitable, consistent boundary conditions at 
$r=0$. Moreover, 
the evolution will be different  for
  different boundary conditions (for a detailed discussion, see \cite{Wald:1980jn, Ishibashi:2003jd,Ishibashi:2004wx,Araneda:2016ecy,rs}).  
 This is what makes the concept of stability 
so subtle: we have to find out if a unique  boundary condition is singled out based on physical arguments. If so, 
determinism is recovered,  dynamics from initial data is uniquely defined and the question of stability makes sense: we have to determine if the LEE 
admit solutions exponentially growing in time \textit{under the physically admissible  boundary condition at $r=0$},   
 while properly  decaying as $r \to \infty$.

We anticipate that there are no unstable odd modes, and that there is  one unstable even mode for every harmonic $\ell=2,3,...$, so we
will concentrate on the even modes from now on. The analysis below follows closely references \cite{Gleiser:2006yz,Dotti:2008ta,ghi}.

When written in terms of $x$, as demanded by Equation (\ref{1+1}), the Zerilli potential in (\ref{1+1}) and (\ref{poten1}) %if and should be added between? the same below
behaves, near the $x=0$ ($r=0$) boundary, as 
$V^{(\ell,m,+)} \simeq -1/(4x^2)$. As a consequence, 
the eigenfunctions (\ref{qh}) of $\mathcal{H}^{(\ell,m,+)}$~in the separable solutions (\ref{sep}) 
behave near $x=0$ as 
\begin{equation}\label{bcs}
\psi^{(\ell,m,+)} = A \cos(\a) \left[ \left(\frac{x}{|M|}\right)^{1/2} + .... \right] + A \sin(\a) \left[ \left(\frac{x}{|M|}\right)^{1/2} 
\ln \left(\frac{x}{|M|}\right)+ .... \right] 
\end{equation}
where $a = A\cos(\a)$ and $b= A \sin(\a)$ are the integration constants and the expressions between square brackets are the leading terms of  
   linearly independent Frobenius series solutions near $x=0$
    (note that these leading terms are independent of the energy eigenvalue $E=\w^2$ and of the spherical harmonic number $\ell$). Thus, 
$\a$ gives the mixture of these two linearly independent solutions that we choose, it parametrizes the different possible boundary conditions 
at the singularity. 
Generic solutions (\ref{bcs}) 
are square integrable in $x$ near $x=0$. A quantum Hamiltonian on the half line $x>0$ like $\mathcal{H}^{(\ell,m,+)}$, for 
which arbitrary eigen-functions are square integrable at the $x=0$ boundary,  
is one of the   cases analyzed in detail in Chapter X in \cite{rs} under the name of ``limit circle at $x=0$'', a terminology that refers 
to the different possible boundary conditions, which are selected by $\a$ in (\ref{bcs}). \\
It was found in \cite{ghi}, that only 
the $\a=0$ solution   is physically acceptable: we have to discard the solution with a logarithm. 
This is so 
because the   first-order contribution of a generic  perturbation  (\ref{bcs}) to the Kretschmann curvature scalar 
invariant $K=C^{\alpha \beta \gamma \delta} C_{\alpha \beta \gamma \delta}$,  
 for a perturbation of the generic form in (\ref{sep}) and  (\ref{bcs}) is (take real part) 
\begin{multline}
\mathcal{K} = \frac{12 M^2}{r^6} - \epsilon A \sin(\a) \left[  \frac{36M^2}{r^7} + \mathcal{O} \left( \frac{\ln(r)}{r^5} \right) \right] S^{(\ell,m)} 
e^{\pm i \w t} \\
 - \epsilon A \cos(\a) \left[ \frac{\ell (\ell^2-1) (\ell+1)(\ell+2)(\ell^2+\ell+4)(\ell^2+\ell-1)}{9M^2r^3} + \mathcal{O}\left(\frac{1}{r^2} 
 \right) \right] S^{(\ell,m)} 
e^{\pm i \w t}
\end{multline}
where $\epsilon$ is the perturbation parameter in (\ref{mpfs}). The above equation is telling us that, no matter 
the $\w$ value (whether real or complex), the perturbation is inconsistent unless we select the boundary 
condition $\a=0$. Otherwise, the perturbation is not initially ($t=0$) uniformly small, that is, no matter how 
small $\e$ is, the perturbation becomes more important than the background as $r \to 0^+$. 
It can be shown that the selection of the $\a=0$ boundary condition in (\ref{bcs}) is preserved by the evolution in (\ref{1+1}), and 
implies that not only $\mathcal{K}$, but 
\textit{all} the curvature scalars made out of the Weyl tensor, the metric tensor and its inverse, 
and the volume form,  share this property with $\mathcal{K}$ \cite{Dotti:2008ta}. Given that it 
guarantees the self-consistency of the linearized treatment, that is, that the smallness of the perturbation at $t=0$ be effectively  
controlled by the perturbation  parameter $\e$, we adopt the boundary condition $\alpha=0$. 
Having made this decision, the evolution of the 1+1 wave equation for the even modes is uniquely defined.

There is, however, a second issue we have to address:  the even potential $V^{(\ell,m,+)}$ in~(\ref{poten1}) has 
a pole at 

\begin{equation}
r_\ell = -\frac{3M}{\lambda}, \; \; \lambda = \frac{(\ell-1)(\ell+2)}{2}
\end{equation}
which is harmless in the BH case, for which $r_\ell<0$, but is in the $r>0$ relevant domain for NSs. 
It is easy to trace back the origin of this singularity: the auxiliary field introduced in~
(\ref{RW2}) is  given in terms of  the  metric perturbation components (\ref{e1})--(\ref{e3}) by 
\begin{equation}
\phi^{(\ell,m,+)} = \frac{r (r-2M)}{(\lambda+1)(\lambda r + 3M)} \left( H_2- r \p_r K \right) + \frac{r}{\lambda+1} K,
\end{equation}
and so it is singular at $r_\ell$ if the metric perturbation is smooth. It is then no surprise that the equation it satisfies be singular 
at this point. This puts a limit to the analogy between the NS linear perturbation problem and its Quantum Mechanics 
analogue (\ref{qh}): for the Quantum Mechanics problem a $C^1$ solution of (\ref{qh}) is acceptable, for the LEE problem 
it is not as, according to (\ref{e1})--(\ref{e3}) and (\ref{RW2}), it gives a discontinuous metric which, as 
explained in \cite{Gleiser:2006yz} does not correspond to a vacuum solution but to a  singular matter source.
In \cite{ghi},  
a $\w=0$ solution $\hat \psi_o$ of the even $\ell=2$ Equation  (\ref{qh}) was found which,  for an specific value $\alpha=\alpha_o$ in (\ref{bcs}), 
satisfies $\psi_o(r_\ell)=0$ and $\hat \psi_o'(r_\ell)=0$. This solutions diverges as $r \to \infty$, but 

\begin{equation}\label{pp}
\tilde \psi_o = \begin{cases} \psi_o &, 0<r<r_\ell\\ 0 &, r>r_\ell \end{cases}
\end{equation}
is  a $C^1$ solution of the equation $\mathcal{H}^{(\ell=2,m,+)} \psi_o= 0$: a marginal mode for a positive spectrum of 
this Hamiltonian under the boundary condition $\a=\a_o$. 
It is  argued in \cite{ghi} that, since $\a_o$ is a marginally stable state, moving $\alpha$ slightly to one or the other side of $\a_o$ would produce 
unstable and stable LEE solutions. However,  although  $\tilde \psi_o$ could be considered a $\w=0$ solution of 
(\ref{qh}), \textit{it does not give} a $\w=0$ solution of the vacuum LEE (\ref{lee}) (see  
the discussion in \cite{Gleiser:2006yz}, Section 7). This is so because $\tilde \psi_o''$ is discontinuous at 
$r=r_\ell$, and so is the metric perturbation (\ref{e1})--(\ref{e3}) made from $\tilde \psi_o$. It can be shown that such a metric 
gives a singular matter distribution (a thin shell of matter) at $r=r_\ell$. The idea in \cite{ghi}  that there is a $\w=0$ mode for an $\alpha_o \neq 0$ 
(recall that $\a=0$ is the physically selected boundary condition in (\ref{bcs})) left unclear the issue of stability of the Schwarzschild NS. 
A few years later, however, an explicit unstable mode for every $\ell$ was found in \cite{Gleiser:2006yz}: 

\begin{equation}\label{us}
\phi_{unst}^{(\ell,m,+)}  = \exp \left[\frac{kt}{-2M}  \right] \;\;
\frac{\left( r - 2 M   \right) ^{k}}{2 \lambda r +6M} \exp \left[\frac{kr}{2M}  \right], 
\end{equation}
  where $k=\frac{ (\ell-1) \ell (\ell+1) (\ell+2)} {6}$ and $\lambda$ was defined in (\ref{RW2a}). 
  This can easily be seen to be a separable perturbation (\ref{sep}) corresponding to the physical 
  boundary condition $\a=0$ in (\ref{bcs}). 
The perturbed metric components for this mode, obtained from  (\ref{e1})--(\ref{e3}), are 
\begin{equation}\label{umc}
\begin{split}
 K(t,r) &= \frac{ (\lambda + 1)(r-2M)^{k }}{6M}\exp \left[\frac{k(t- r)}{-2M}  \right], \\
 H_1(r,t)  &= -\frac{\lambda (\lambda+1)
   [2(\lambda+1) r-6M] \; r  (r-2M)^{k-1}}{36M^2}  \exp \left[\frac{k(t- r)}{-2M}  \right] =-H_2(t,r)
   \end{split}
   \end{equation}
   
Note that the metric components  (\ref{e1})--(\ref{e3}) for (\ref{umc}) are: (i) smooth 
for $0<r<\infty$, (ii) vanishing at $r=0$, 
 exponentially decaying  as $r \to \infty$ and that, 
since they grow exponentially in time, they signal  an instability.

The effect of these unstable perturbations on the spacetime geometry is rather subtle and deserve a few comments: 
the unstable solution (\ref{us}) was found by extrapolating to generic $\ell$ the results obtained by a shooting 
approach on Equation (\ref{qh}) for $\ell=2,3$, from where $k=\frac{ (\ell-1) \ell (\ell+1) (\ell+2)}{6}$ 
was correctly guessed. 
It was then pointed out in \cite{Cardoso:2006bv} that they correspond to Chandrasekhar \textit{algebraically special modes} 
\cite{ch1,ch2}, which had not been paid much attention because they are exponentially growing with $r$ when $M>0$, and 
thus irrelevant as BH perturbations (see (\ref{us})). In \cite{Cardoso:2006bv},  this observation was used to 
prove the instability of the Schwarzschild de Sitter spacetime, and correctly guess that the algebraically special modes 
would play a role in the instability of the super-extreme  Reissner-Nordström solution.
They do not, however, play a role in the instability of the Kerr NS, as they do not satisfy suitable boundary conditions in this case. 
One of the paritcularities of these modes is that the first order perturbation of the algebraic curvature scalars made out of the Weyl tensor 
vanish. The effect of the unstable NS perturbation on the geometry can be seen in \textit{differential} 
scalar curvature scalars. As an example, it was found in \cite{Gleiser:2006yz} that 
\begin{equation}\label{este}
R_{\a \beta \gamma \delta;\e} R^{\a \beta \gamma \delta;\e} = \frac{720 M^2 (r + 2 |M|)}{r^9}\\ - \epsilon \; \frac{5 (\ell+2)! (\ell (\ell+1) r +6|M|)
(r + 2 |M|)^{2k |M|} }{(\ell-2)! r^8}\;  e^{k(t-r)} \;S^{\ell m}(\theta,\phi)
\end{equation}

Note again from this equation that, 
 at $t=0$, the  perturbation can be made \textit{uniformly}  small  in the $0<r<\infty$ interval (relative to the unperturbed value)  
 by choosing $\e$ small enough, as it has an $r^{-8}$ pole 
whereas the unperturbed scalar field behaves as $r^{-9}$ as $r \to 0^+$. The instability is signaled by 
the fact that the perturbation grows exponentially with time. As a consequence, as happens with any unstable solution, the linearized approximation 
breaks down soon.  Another first order effect of the unstable perturbation is the splitting of one pair 
of degenerate PNDs (solutions of (\ref{pnd-d}), recall that the background NS is Petrov type D). This is proved in \cite{Araneda2015}, where it was 
found that, for the null tetrad

\begin{equation}\label{snt}
\begin{split}
l^a \p_a &= (2f)^{-1/2} \p_t - (f/2)^{1/2} \p_r\\
n^a \p_a &= (2f)^{-1/2} \p_t + (f/2)^{1/2} \p_r\\
m^a \p_a&= r^{-1} \p_\theta - i (r \sin \theta)^{-1} \p_\phi
\end{split}
\end{equation}
at leading order $n^a$ does not split, whereas $l^a$ splits into two solutions of  Equation (\ref{pnd-I}):
\begin{equation} \label{ls}
l^a_{\pm}(\epsilon):=l^a\pm\epsilon^{1/2}\left[\overline{\left(\sqrt{-\frac{\delta\Psi_4}
{6\Psi_2}}\right)}\bar{m}^a+\sqrt{-\frac{\delta\Psi_4}{6\Psi_2}}m^a  \right] 
\end{equation}
where the spin-weighted Weyl scalars were defined in (\ref{np1}), the background value is $\Psi_2= -M/r^3$, and 
\begin{equation}
\delta \Psi_4 = \frac{6k}{M^2}\; 
(r-2M)^{k-1} \left[(\ell+2)(\ell-1) + \frac{6M}{r} \right]  \; \exp \left( \frac{k(r-t)}{2M} \right) \; S^{(\ell,m)}(\theta,\phi).
\end{equation}

 The non-analytical character of the splitting (as a function of the perturbation parameter $\e$), discussed in 
some detail in \cite{cherubini}, can be avoided by a re-parametrization of the familiy of 
solutions $g_{ab}(\epsilon)$ in (\ref{mpfs}).

The evolution of  gravitational perturbations from initial data on the NS Schwarzschild background 
 cannot be afforded with Equation (\ref{1+1}), since 
$\phi^{(\ell,m,+)}$ is, as we commented, ill defined, carrying a built-in pole at $r=r_\ell$. We need to switch to a 
well defined variable in the even sector carrying the same information as the Zerilli field $\phi^{(\ell,m,+)}$. This is done 
in detail in the paper \cite{Dotti:2008ta}, where the initial value formulation for the even LEE is dealt with, and 
it is 
shown that \textit{generic} initial perturbation data  have a projection on the unstable modes  and thus develop 
an instability.

The technique used in \cite{Dotti:2008ta,Gleiser:2006yz}  was then applied in \cite{Dotti:2010uc}  to prove that the super-extreme Reissner-Nordstöm 
solution, a NS within the Kerr--Newman family of solutions of the Einstein--Maxwell equations, is also unstable.
 The instability is found again in the even modes, one for every harmonic. 
The treatment, however, is more complex, because of the larger number of degrees of freedom.

\subsection{Instability of the Kerr Naked Singularity}\label{Siks}

The gravitational instability of the Kerr NS was proved in \cite{dgr,Dotti:2011eq}. In \cite{dgr}, it was shown  show that there
 are solutions of the $s=-2, m=0$ Teukolsky 
Equations (\ref{teukofield})--(\ref{tr})  with 
\begin{equation} \label{wika}
\w= ik/a, k, \;\;\; k>0,
\end{equation}
that is, exponentially growing in time. 
 Note the following requirements on admissible  solutions of (\ref{ta}) and (\ref{tr}): 
 \vspace{+3pt}
 \begin{enumerate}%[leftmargin=27pt,labelsep=7pt]
\item[(i)]   $S^{(\ell,m)}_{\w,s}(\theta) e^{im\phi}$ has to  be smooth on the sphere (as we mentioned, these  are called 
\textit{spin weighted spheroidal harmonics (SWSH)}). This  gives a 
 a discrete spectrum of $\w-$dependent eigenvalues $E=E_\ell(\w)$, $\ell=0,1,2,...$;  \\
\item[(ii)] The radial functions $R$, which  have domain  $-\infty < r <\infty$ for the 
 $a>M>0$ Kerr NS, must decay as $r \to  \pm \infty$.
  We will show below that this also forces the spectrum of radial $E$'s in (\ref{tr}) to  be discrete.
  We will call   $\e_o(\w)$ the  fundamental state for a given admissible $\w$. \\
\item[(iii)]  $E$ is the same in (\ref{ta}) and (\ref{tr}).\\ 
 The strategy of the proof of instability is to show that, for $s=-2$ and $m=0$, 
 there is a value $k^*>0$ of $k$ such that $E_{\ell=2}(\w=ik^*/a)=\e_o(\w=ik^*/a)$. This  implies that the 
 Teukolsky system of equations 
 (\ref{teukofield})--(\ref{tr}) admits an $\ell=2,m=0$ gravitational ($s=-2$) mode satisfying appropriate boundary conditions,  and 
 exponentially growing in time as $e^{k^*t}$. The proof  below is reproduced  from \cite{dgr}.
\end{enumerate}
 
% From now on we simplify the notation $E_{\ell=2}(\w=ik/a), \e_o(\w=ik/a)$ to $E_\ell(k)$ and $\e_o(k)$, respectively, $k$ a positive real number. 
We proceed in three steps: first we gather information on the $E_\ell(k)$ spectrum of SWSH, 
 then we study the behavior of $\e_o(k)$, the fundamental state of the radial Equation 
(\ref{tr}), as a function of $k$. Finally, we prove that, for any $\ell$ there is a $k^*>0$ such that 
$\e_o(k^*)=E_\ell(k^*)$.

  For complex $\w$ near $\w=0$, a Taylor expansion for $E_{\ell}$ for regular solutions of the angular 
  Equation (\ref{ta})  
 up to order $(a \w)^6$ can be found in the literature (\cite{seidel,bertilong} and references therein). Asymptotic expansions
 in the limits where  $\w \to \infty$ and $\w \to +i \infty$ are also available  \cite{lnp,brw,bertishort,bertilong}.
For the axial case $m=0$, and for gravitational perturbations $s= \pm 2$,
  the $E_\ell$ angular eigenvalue for  large purely imaginary $\w$ as in (\ref{wika}) 
  behaves as \cite{bertilong}
   \begin{equation} \label{arrivaberti}
 E_{\ell}(\w=ik/a) = (2 \ell -3) k + {\cal O} (k^0), \;\;\; \text{ as }  k \to \infty
 \end{equation}
 
 Here we adopt  the (non generalized) convention 
 that  the $\ell$ labeling is  such that the $(\ell,m,s)$ spin weighted spheroidal harmonic $S$ 
reduces to the corresponding spin weighted {\em spherical} harmonic in the $a \w=0$ limit \cite{bertilong}, for which 
\begin{equation} \label{abajoberti}
E_{\ell}(0) = \ell(\ell+1) 
\end{equation}

Under this convention, the relevant modes for gravitational perturbations are $\ell=2,3,4,...$  
In \cite{dgr}, the low frequency approximations  in \cite{seidel,bertilong} as well as the 
asymptotic formula (\ref{arrivaberti}) were numerically checked by finding Frobenius expansions 
of the solutions of (\ref{ta}) 
centered at the poles $\cos \theta \equiv x= \pm 1$, and having they match (together with its derivative) at $x=0$ 
(this produces an equation for the eigenvalue $E$ that was solved numerically). 
The agreement found with (\ref{abajoberti}) and the $\w \simeq 0$ Taylor expansions in the literature 
is excellent. Equations  (\ref{arrivaberti}) and (\ref{abajoberti}) give the information 
we need about the angular eigenvalues $E_\ell(\w=ik/a)$. We now comment on the strategy to learn 
about the behavior of the fundamental eigenvalue $\e_o(\w=ik/a)$ of the radial Equation (\ref{tr}). 

Introducing 
\begin{equation} \label{rs}
x = \ln \left( \frac{r-M+\sqrt{r^2-2Mr+a^2}}{M} \right)  \simeq \begin{cases} \ln \left(\frac{2 r}{M} \right) & r \to \infty \\
\ln \left( \frac{a^2-M^2}{2 M |r| } \right) & r \to - \infty \end{cases}
\end{equation}
which grows monotonically with $r$ and has a simple inverse:
\begin{equation} \label{irs}
r = \frac{M \exp(x)}{2} + M + \frac{M^2-a^2}{2M \exp(x)},
\end{equation}

 Equation (\ref{tr})  can be put in the form  of a %please consider this change [AGREED]
 a Schr\"odinger equation
 \begin{equation} \label{scho}
{\cal H} \psi := - d^2\psi /dx^2 + V \psi = -E \psi,
\end{equation}
with a  potential (recall that we fixed $s=-2, \w=ik/a, m=0$) 
\begin{multline} \label{v}
V = \left[\frac{r^4+ a^2 r^2 +2 a^2 M r}{a^2(r^2-2Mr+a^2)} \right]  \; k^2 +
 4 \left[ \frac{-r^3 + 3 M r^2-  a^2r-a^2 M}{a(r^2-2Mr+a^2)} \right]  \; k
   + \left[ \frac{r^2-2Mr+15 M^2-14a^2}{4(r^2-2Mr+a^2)}
 \right]\\ =: k^2 V_2 + k V_1 + V_o
\end{multline}
where $r$ given in (\ref{rs}). 
 $V$ is smooth everywhere because $a>M>0$, and  is bounded from below for every $k \geq 0$. 
 There is a 
 subtlety here: the   minimum of $V$ is not 
  continuous, as a function of $k$, at $k=0$:
  
 \begin{eqnarray} \label{vm}
 \min \{ V(r,k=0), r \in {\mathbb R} \} & = & -7/2 \\ \nonumber
 \lim_{k \to 0^+}  \min \{ V(r,k), r \in {\mathbb R} \} & = & -15/4
 \end{eqnarray}
 
The operator  ${\cal H}$ defined in (\ref{scho}) is self-adjoint in the Hilbert space of square integrable functions of $x$ (note 
that $-\infty<x<\infty$) and,   
  since $V$ is  bounded from below and
  
 \begin{equation} \label{a1}
 V \sim  \left( \frac{Mk e^{x}}{2a} \right) ^2 ,  |x| \to \infty,
 \end{equation}
 the spectrum of  ${\cal H}$ is, as we anticipated,  fully discrete, and has a lower bound $E=\e_o(\w=ik/a)$. 

\begin{comment}
The square integrable (in $x$) eigenfunctions of ${\cal H}$ behave
as

\begin{equation} \label{ae}
\psi \sim \begin{cases} e^{- \frac{rk}{a}} \left( \frac{M}{r} \right) ^ {\frac{2kM}{a}-3} \left( 1 + {\cal O} (M/r) \right)  & , r \to \infty \\
                    e^{ \frac{rk}{a}} \left( \frac{M}{r} \right) ^ {1-\frac{2kM}{a}} \left( 1 + {\cal O} (M/r) \right) & , r \to -\infty  \end{cases}
\end{equation}

with sub-leading terms that depend on the eigenvalue.\\

\end{comment}
To obtain information about the fundamental energy of the radial Hamiltonian (\ref{scho}), we  analyze
the potential (\ref{v}). The numerator of $V_2$,
$r (r^3+ a^2 r  + 2a^2M )$, is negative in some interval to the left of zero, and, since
 $V_1(0) = -4a^2M <0$, by continuity $V_1$   must also be negative in some neighborhood  of $r=0$.
 It follows that  there is  an interval $s_1(M)< r < s_2(M)<0$ where both
 $V_1$ and $V_2$ are negative.
Let $\psi$ be a
smooth normalized test function
supported on
$r(x) \in (s_1(M),s_2(M))$. Then 
  $\psi$ has  compact support away from $r=0$, and 
  \begin{equation} \label{test1}
\langle \psi | {\cal H} | \psi \rangle =   \langle \psi | - ( \partial / \partial x ) ^2 |
\psi \rangle +
\sum_{j=0}^{2} k^j \langle \psi |V_j |\psi \rangle
\end{equation}
with
\begin{equation} \label{test2}
\langle \psi |V_j |\psi \rangle =
 \int_{s_1(M)}^{s_2(M)} | \psi |^2 \; V_j \; \frac{dr}{\sqrt{\Delta}} < 0
\text{ for } j=1,2.
\end{equation}

Since $\langle \psi |V_1 |\psi \rangle$  is negative, there is a $k_c$ such that,
for $k>k_c$,
$ \langle \psi | - ( \partial / \partial x) ^2 |
\psi \rangle +
\langle \psi |V_0 |\psi \rangle + k \langle \psi |V_1 |\psi \rangle $ is  negative
and so $\langle \psi | {\cal H} | \psi \rangle  <  k^2  \langle \psi |V_2 |\psi \rangle$.
This implies that if $-\epsilon_o(\w=ik/a)$
is the lowest  eigenvalue of ${\cal H}$ then

\begin{equation} \label{br}
 -\epsilon_o(\w=ik/a) \leq \;  \langle \psi | {\cal H} | \psi \rangle  <  k^2  \langle \psi |V_2 |\psi \rangle  < 0 \, , \; \; \text{ if } k > k_c.
\end{equation}

From (\ref{vm}) and the above equation follows that 
\begin{equation} \label{radiale}
\epsilon_o(\w=ik/a)\mid_{k=0^+}   <  \frac{15}{4}, \;\;\; \text{ whereas } \;\;\; 
 \epsilon_o(\w=ik/a) > | \langle \psi |V_2 |\psi \rangle | k^2 , \; \; k > k_c.
\end{equation}

On the other hand, using (\ref{arrivaberti}), (\ref{abajoberti}) and (\ref{radiale}) it is clear that for every 
$\ell=2,3,...$, the curves $\epsilon_o(\w=ik/a)$ and  $E_{\ell}(\w=ik/a)$ intersect at some
$k_{\ell} > 0$. This behavior  was checked numerically in a specific case in  \cite{dgr}. The 
  results, shown in  Figure~\ref{F1}, agree with previous estimations 
 in \cite{doglepu}.
\begin{figure}[h]
\includegraphics[width=10.5 cm]{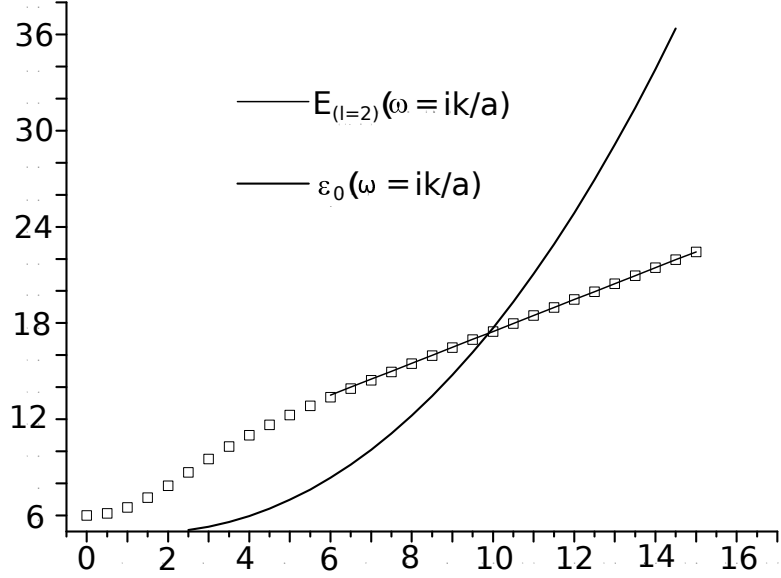}
\caption{Intersection of the numerically generated %if copyright permission is needed for this figure? please confirm [EDITED 
%A NEW EPS FILE TO MAKE IT DIFFERENT FROM THE ONE IN CQG AND SAVE PERMISSION TIMES] 
curves $E_{(\ell=2)}(\w=ik/a)$ and $\epsilon_o(\w=ik/a)$  for $a/M=1.4$. 
Note the agreement with the results in \cite{doglepu}, according to which  for $a=1.4$ the intersection occurs at $k \simeq
7.07 \times 1.4  = 9898$.}
\end{figure}

We close this section with two remarks. The first one  concerns the  behavior of 
 the perturbation
at the ring singularity. Contrary to what happens for the static 
 NSs such as the negative mass Schwarzschild solution, or the super-extreme Reisner-Nordström solution, 
$r=0$ is {\em not} a singular point of the radial equation of the Kerr NS: the potential (\ref{v}) is indeed 
 smooth everywhere. In fact,
$r=0$ is neither a boundary point,
nor a singular point of the second order radial Equation (\ref{scho}), for which the domain is $-\infty < r < \infty$. 
Imposing  any conditions at $r=0$,
besides requiring that the perturbations vanishes at
 $r = \pm \infty$,  leads to an unnaturally
 over-determined problem. There is no visible effect of the ring singularity at this level. 
The second remark is that 
 unstable perturbations are
 not restricted to the fundamental radial mode since, for
  large $k$, the potential $V$ in (\ref{scho}) has a deep minimum
   that can be approximated by a harmonic oscillator potential of depth of order $k^2$, and
   strength also of order $k^2$, so that the level spacing near the ground state of  ${\cal H}$ is
   of order $k$. Thus, the negative   $-\epsilon_n(k)$ of the lowest radial eigenvalues
 grow also quadratically in $k$  for large $k$, and  intersect the angular eigenvalue curves for
 sufficiently large values of $k$.

\subsection{Instability of the Kerr Black Hole  Region beyond the Cauchy Horizon}\label{Sibch}

In Section \ref{Sisns}, we commented that the Reissner-Nordström NS and the 
 region beyond the Cauchy (that is, the inner) horizon of a Reissner-Nordström BH are unstable, 
and that the proof on instability follows similar steps as the proof of instablility of the Schwarzschild NS 
(see \cite{Dotti:2010uc}) for details). 
In what follows we show how the proof  of instability of the Kerr NS above can be adapted to 
prove  the instability of the $r<r_-$ region of a Kerr BH. The extreme $a=M$  and
 sub-extreme $a<M$ cases require separate treatments. 
 
\vspace{+3pt}
 \subsubsection{Extreme case (a=M) inner region:}%is the bold necessary? the same below [SUPPRESSED]
 
 The solution
 of the equation $dx/dr = 1/\sqrt{\Delta}$ in the interior region  $r < r_- = r_+ = M$ is
  \begin{equation} \label{rse}
x =
-\ln \left( \frac{M-r}{M}  \right), \;\; r < r_-,
\end{equation}
its inverse is 
\begin{equation} \label{irse}
r = M ( 1- e^{-x}) , \;\;  -\infty < x < \infty.
\end{equation}

Using  the
  integration factor $\Delta^{3/4}$ as before, we are led back to (\ref{scho}) and (\ref{v}), with
 $r$ given in (\ref{irse}).
  Note that
   \begin{equation} \label{av2}
V \sim \begin{cases}  4k^2 \exp(2x)  & , x \to \infty \\
                k^2 \exp(-2x)     & , x \to -\infty , \end{cases}
\end{equation}
then the   spectrum of the self-adjoint
  operator ${\cal H}$  is again fully discrete and has a lower bound. The eigenfunctions behave as
   \begin{equation} \label{aee}
\psi \sim \begin{cases} \left(\frac{M}{M-r} \right)^ {2k} \; \exp \left[-2k \left(\frac{M}{M-r} \right) \right]
 \left( 1 + {\cal O} (\frac{M-r}{r}) \right)  & , r \to M^- \\
                    \frac{M}{r}^{1-2k} \; e^{ \frac{rk}{M}}  \left( 1 + {\cal O} (M/r) \right) & , r \to -\infty  \end{cases}
\end{equation}

The argument of instability that we used 
in for the Kerr NS goes through in this case without modifications, since the test function in (\ref{test1}) is supported in the 
$r<0$ region,
and thus can be used again in this case to obtain the bound (\ref{br}).
The radial decay (\ref{aee}) guarantees that corrections to relevant quantities vanish in these limits.
The considerations
about  the perturbation behavior near the ring singularity for the Kerr NS case apply
verbatim   to this case, and also to the sub-extreme case below. \\

\subsubsection{Sub-extreme case ($a < M$) inner region:}

The sub-extreme case introduces some subtleties: requiring $dx/dr = 1/\sqrt{\Delta}$ for  $r< r_-$~gives

\begin{equation} \label{urs}
\begin{split}
x &= \ln \left( \frac{M-r-\sqrt{r^2-2Mr+a^2}}{M} \right)\\
 &= \ln \left( \frac{r_- + r_+-2r-2 \sqrt{(r_+-r)(r_--r)}}{r_-+r_+} \right), \;\; r< r_- ,
 \end{split}
 \end{equation}
 and the radial equation reduces to (\ref{scho}), (\ref{v}) with
$r$ the inverse of (\ref{urs}). 

Since
\begin{equation}
-\infty < x < x_- :=  \ln \left( \frac{r_+ - r_-}{r_++r_-} \right) ,
\end{equation}

Equation  (\ref{scho}) is, in this case,  a
Schr\"odinger  equation  \textit{on a  half axis}, with
  potential
     diverging  as 
          $$V \sim [k (M^2-a^2)/(2aM)]^2~\exp(-2 x)$$ for  $x \to -\infty$.  For  $x \to x_-{}^-$, 
     
%\begin{adjustwidth}{-\extralength}{0cm}
%\centering %% If there is a figure in wide page, please release command \centering
 \begin{equation}  \label{sqp}
V \sim \left[ -\frac{1}{4} + \frac{4M^2 \left( M-\sqrt{M^2-a^2} \right) ^2}{M^2-a^2} \left(
\frac{k}{a} - \frac{\sqrt{M^2-a^2}}{M \left( M - \sqrt{M^2-a^2} \right)} \right)^2 \right] \frac{1}{(x_--x)^2} + \cdots =: 
\frac{\nu(k)^2-\frac{1}{4}}{(x_--x)^2}+ \cdots ,
\end{equation}
%\end{adjustwidth}
with  $\nu(k) > 0$.
 The eigenfunctions of ${\cal H}$  behave near $x=x_-$ as 
 \begin{equation} 
\psi \sim   A \cos (\a)  \left[(x_--x)^{\frac{1}{2}+ \nu} + ... \right] + A \sin(\a)  \left[ (x_--x)^{\frac{1}{2}- \nu} + ... \right] ,  \label{zero}
\end{equation}
 (the $E$ eigenvalue appears  in sub-leading terms).  
  If  $\nu>1$,
for  generic $E$ these are   not square integrable near zero unless we choose  $\a=0$. A case like this referred to as ``limit point at $x=x_-$''  
in the classification introduced in chapter X in \cite{rs} (recall that in the case of the Schwarzschild NS we had 
 \textit{limit circle case}). 
 In this case we need to choose $\a=0$ to define 
  a space of functions
where ${\cal H}$ is self-adjoint. This selects a discrete set of possible $E$ values as the spectrum of ${\cal H}$. 
On the other hand, if $\nu<1$, any eigenfunction (\ref{zero}) will be  square integrable
near the boundary  $x =x_-$, that is, $V$ is limit circle at $x_-$ if $\nu<1$.  
In this case, a choice of boundary condition, that is,  a fixed value of $\a$ in (\ref{zero}), needs to be made to define the domain of 
 allowed perturbations. This is entirely analogous to the situation 
 we found when studying Schwarzschild (see the discussion following Equation (\ref{bcs})). 
In this case,  however, it is easy to  prove that, 
\textit{regardless  our choice for $\a$} in (\ref{zero}) there will be unstable modes. 
This is so because  functions of compact support away from $x_-$ 
belong to  the domain, for any $\a$, and the test function used in
(\ref{test1})--(\ref{test2}) is of compact support, so the proof of existence of an unstable mode 
goes through for any $\a$.

\section{Nonmodal Stability of the Schwarzschild Black Hole}\label{nms}

Prior to \cite{Dotti:2013uxa} all notions of stability of the exterior region  of a Schwarzschild BH were concerned 
with finding bounds to the Regge-Wheeler and Zerilli fields $\phi^{(\ell,m,\pm)}$. These fields 
are  defined in the two dimensional $(t,r)$ \textit{orbit space} 
and parametrize time dependent metric perturbations in the Regge-Wheeler (RW) gauge as shown in the set of equations,~(\ref{gp})--(\ref{poten1}), that we will write more concisely as

\begin{equation} \label{dec}
{}^{RW}h_{\a \beta}= \sum_{(\ell,m,p=\pm)} h_{\a \beta}^{(\ell,m,p)},
\end{equation}
where 
\begin{equation} \label{rws}
{}^{RW}h_{\a \beta}^{(\ell,m,p=\pm)}=\mathcal{D}_{\a \beta}^{(\ell,m,p=\pm)}
\left[\phi^{(\ell,m,\pm)},S_{(\ell,m)}\right]
\end{equation}
and $\mathcal{D}_{\a \beta}^{(\ell,m,p=\pm)}$ is a second order differential operator. 

 Some key results were:
 
\begin{itemize}
\item In \cite{Regge:1957td}, it was shown that  separable solutions $\phi^{(\ell,m,-)} = \Re\; e^{i \w t} \psi^{(\ell,m,-)}(r)$ 
that do not diverge as $r \to \infty$ require $\w \in \mathbb{R}$, ruling out exponentially growing solutions in the odd sector. 
\item In \cite{Zerilli:1970se}, it was shown that  separable solutions $\phi^{(\ell,m,+)} = \Re \; e^{i \w t} \psi^{(\ell,m,+)}(r)$ 
that do not diverge as $r \to \infty$ require $\w \in \mathbb{R}$, ruling out exponentially growing solutions in the even sector. 
\item In \cite{Price:1971fb}, it was shown that, for large $t$ and  fixed $r$,   $\phi^{(\ell,m,\pm)}(t,r)$ decays  as

\begin{equation}\label{pt}
\phi^{(\ell,m,\pm)}(t,r_o) \sim  t^{-(2\ell+2)},
 \end{equation}
 
an effect known as ``Price tails''.
\item In \cite{wald1}, the conserved energy

\begin{equation}
\int_{2M}^{\infty} \left[(\p_t \phi^{(\ell,m,\pm)})^2 + (\p_x \phi^{(\ell,m,\pm)})^2 +  
f V^{(\ell,m,\pm)} {\phi^{(\ell,m,\pm)}}^2
\right] dx
\end{equation}

was used to rule out uniform exponential growth in time.
\item Furthermore, in \cite{wald1}, a pointwise bound 
\begin{equation} \label{modal-b}
|\phi^{(\ell,m,\pm)}(t,r)| < C_{(\ell,m)}^{\pm}, \;\; r>2M, \text{ all } t,
\end{equation}
was proved, where the constants $C_{(\ell,m)}^{\pm}$ are given in terms of the initial data  

\begin{equation} \label{modal-c}
(\phi^{(\ell,m,\pm)}(t_o,r), \p_t \phi^{(\ell,m,\pm)}(t_o,r)).
\end{equation}
\end{itemize}

 To understand the limitations of these results it is important to keep in mind that 
  the $ \phi^{(\ell,m,\pm)}$ are  an infinite set of  fields defined on the $(t,r)$ orbit space, \textit{whose   
first and second order derivatives} 
 enter the terms in the series (\ref{dec})  through (\ref{rws}), \textit{together with the sheperical harmonics and their derivatives}. 
 Two extra derivatives are then required  to calculate the perturbed Riemann tensor from the metric perturbation, as  a  first step  to measure  the 
effects of the perturbation on the curvature. Thus,   the relation of the   $\phi^{(\ell,m,\pm)}$
 to geometrically meaningful
quantities is remote, and the usefulness 
  of the bounds (\ref{modal-b}) to measure the strength of the perturbation   is  not obvious at all. 

  Even if we knew the impact of these bounds on the components of ${}^{RW}h_{\a \beta}$, we would face 
  the unavoidable problem of the lack of a natural measure of the ``size'' of tensor fields on a Lorentzian manifold. 
  On a Riemannian manifold, where the metric is positive definite, the pointwise size of a $(0,2)$ tensor $h_{\a \beta}$, 
could be    measured by $\sqrt{g^{\a \gamma} g^{\beta \delta} h_{\a \beta} h_{\gamma \delta }}$, which is positive if $h_{\a \beta }$ is nonzero, and an 
  $L^2$ norm of the field  given by $\left( \int_M g^{\a \gamma} g^{\beta \delta } h_{\a \beta } h_{\gamma \delta } \right)^{1/2}$. None of 
  these notions is available for a Lorentzian $g_{\a \beta }$.
  
  Besides the problem, inherent to Lorentzian geometry, of measuring the ``size'' of tensors, there is the often overlooked 
  fact that controlling the size of time dependent series terms of a quantity does not entirely control the quantity itself. 
  This is what  led to the notion of \textit{nonmodal stability} in fluid dynamics, 
   where the limitations of the  mode analysis 
were realized   in 
 experiments involving  shear flows  bounded by walls \cite{schmid}. In this case,   the 
linearized Navier--Stokes operator is non normal, so their eigenfunctions are non orthogonal. As a consequence, even 
if the individual modes decay as $e^{i w t}, \Im (w) >0$---a condition that assures large $t$ stability---there may be important transient 
growths \cite{schmid}. Take, for example, the following toy model (from Section 2.3 in \cite{schmid})  %if this refers to section in ref. 36? YES, FIXED
of a system with two degrees of freedoms: $\vec{v} \in 
\mathbb{R}^2$ obeying the equation $d \vec{v}/dt = -A \vec{v}$, with $[A,A^T] \neq 0$ a matrix with (non orthogonal!) eigenvectors 
$\vec{e}_1, \vec{e}_2$, say, 

\begin{equation}
A = \left( \begin{array}{cc} 1 & \g \\ 0 & 2 \end{array} \right), \;\;\;  
\vec{e}_1= \left( \begin{array}{c} 1 \\ 0 \end{array} \right),  \;\;\;  
\vec{e}_2= \left( \begin{array}{c} 1 \\ \g^{-1} \end{array} \right)
\end{equation}

Consider the case $\g >>1$. Note that, although $\w_1=i, \w_2=2i$, that is, normal modes decay exponentially,  if 
 $\vec{v}(0)=\a (\vec{e}_1-\vec{e}_2)$, then $\vec{v}(t)=\a \vec{e}_1 e^{-t}- \a \vec{e}_2 e^{-2t}$ 
reaches a maximum norm $||\vec{v}|| \simeq (\g/4) ||\vec{v}(0)||$ at a finite time before decaying to zero. 
The operators involved in the LEE are normal, so the above situation of non orthogonal eigenfunctions 
does not arise. However, it is easy to construct examples of, e.g., time dependent 
scalars on the sphere, say 

\begin{equation}
Z(t,\theta,\phi) = \sum_{\ell,m} z_{(\ell,m)}(t) S^{(\ell,m)}(\theta,\phi)
\end{equation}
 where 
the $z_{(\ell,m)}(t)$ oscillate  (as the pure modes (\ref{sep}) were shown to do) and yet the function 
$S(t,\theta,\phi)$ develops arbitrarily high localized transient growths, or even grows without bound  as $t \to \infty$ in 
continuously narrowing areas. This is so because  (recall 
that our $S^{(\ell,m)}(\theta,\phi)$ are an orthonormal basis of \textit{real} spherical harmonics on the unit sphere $S^2$)

\begin{equation}
z_{(\ell,m)}(t) = \frac{1}{4 \pi} \int_0^\pi  \sin  \theta \; d \theta \int_{0}^{2 \pi} Z(t,\theta,\phi)  S^{(\ell,m)}(\theta,\phi)\;  d \phi, 
\end{equation}
and the integral above can be kept bounded in $t$ while at $Z(t,\theta,\phi)$ grows high in narrowing areas. 
In fact, this was precisely the motivation in \cite{wald1}: to ``undo'' the  separation of variables~in 

\begin{equation}\label{ft}
\phi^{(\ell,m,\pm)}(t,r) = \int d\w \; \psi^{(\ell,m,\pm)}(x(r)) \;e^{i \w t}
\end{equation}
and show that the sum (\ref{ft}) of oscillating modes does not allow unbounded  transient growths in $\phi^{(\ell,m,\pm)}(t,r)$, and that 
we can place pointwise bounds based on the initial data of the form (\ref{modal-b}) and (\ref{modal-c}). This could be done 
exploiting the form of the partial differential Equation (\ref{1+1}) obeyed by the fields $\phi^{(\ell,m,\pm)}$. 
Although finding an exponentially growing mode is a definite signal of instability, as we can see, there are different possible degrees of stability 
when all modes are oscillating (real $\w$'s). The symmetries of the background geometry is what allowed us to decompose metric perturbations as 
in (\ref{rws}) and (\ref{sep}). The results in \cite{wald1} can be regarded as a way to  ``undo''  the $t-r$ variable separation (\ref{sep}): they prove 
 (\ref{modal-b}) and (\ref{modal-c})  for \textit{arbitrary} (that is, non necessarily separable) solutions of the 1+1 wave Equations (\ref{1+1}). 
A natural question after the work \cite{wald1} is: Can we also ``undo'' the $(t,r)-(\theta,\phi)$ separation of variables? Can we 
find bounds for arbitrary 
solutions of the LEE (\ref{lee})?
The question is very tricky since it omits a difficult issue: which \textit{spacetime scalar field} should we try to place bounds on?

  The strategy in \cite{Dotti:2013uxa} was to measure the pointwise intensity of a perturbation by its effect on 
  curvature related scalar fields (CSs, for short, not to be confused with the Newman-Penrose \textit{spin weighted} scalars (\ref{np1}) and (\ref{np2}), which 
  depend point by point on a selected null tetrad). These scalars are full contractions of the Weyl tensor $ C_{\a \beta \g \delta}$ 
  (which equals the Riemann tensor in vacuum), 
  its covariant derivatives, the metric and its inverse, and the volume form $\varepsilon_{\a \beta \mu \nu}$. Some examples are
   \begin{equation} \label{curvaturescalars}
 \begin{split}
&Q_- = \tfrac{1}{96} C^{\a \beta \g \delta} \varepsilon_{\a \beta \mu \nu}  C^{\mu \nu}{}_{\g \delta}, \;\;\;\;\;
 Q_+ = \tfrac{1}{48} C^{\a \beta \g \delta}   C_{\a \beta \g \delta} , \\
&X = \tfrac{1}{720}  \left( \nabla_{\e} C_{\a \beta \g \delta} \right)  \left( \nabla^{\e} C^{\a \beta \g \delta} \right).
\end{split}
\end{equation} 

Working with scalar fields avoids the issue discussed above 
of ``measuring the size of tensors'' when the background metric is Lorentzian. There is, however, the misconception 
that the first order variation of any CS is gauge invariant (that is, coordinate independent). This is not the case: under the transformation 
(\ref{gt2}), the first order perturbation of a CS field $Z$ changes as $\delta Z \to \delta Z' =\delta Z + \pounds_{\xi}  Z$, so $\delta Z$ is gauge invariant 
if and only if the background CS vanishes: $Z=0$. As an example, in the Schwarzschid de Sitter (SdS) background, the CSs    
(\ref{curvaturescalars}) take the values
\begin{equation}\label{bvcs}
{Q_-}_{SdS} = 0, \;\;\; {Q_{+}}_{SdS} = \frac{M^2}{r^6}, \;\;\; X_{S(A)dS} = \frac{M^2}{r^9} (r-2M) - \frac{\Lambda M^2}{3 r^6}, 
\end{equation}
so in the linearized theory 
only $\delta Q_-$is a gauge invariant quantity. However, \textit{combinations} of first order variations of CSs whose background values 
are nonzero can be gauge invariant. As an example, the field  
\begin{equation} \label{Gp}
G_+ = (9M-4r+\Lambda r^3) \dot Q_+ + 3 r^3 \dot X
\end{equation}
is gauge invariant since, under a gauge transformation along $\zeta^{\a}$,
\begin{align}\begin{split}
G_+ \to &G_+ +  (9M-4r+\Lambda r^3)\; \pounds_{\zeta}  {Q_+}_{SdS}   + 3 r^3 \; \pounds_{\zeta}  X_{SdS}  \\
& =   G_+ +  (9M-4r+\Lambda r^3) \;  \zeta^r \p_r  {Q_+}_{SdS}   + 3 r^3  \;  \zeta^r \p_r  X_{SdS}  \\
& = G_+ \end{split}
\end{align}

In \cite{Dotti:2013uxa}, the fields $G_-=Q_-$ and $G_+$ were proposed to measure the strength of perturbations. 
It was shown that: 

\begin{enumerate}
\item  $G_-$ depends only on the odd piece $h_{\a \beta}^-=\sum_{\ell,m} h_{\a \beta}^{(\ell,m,-)}$ of the perturbation, 
whereas  $G_+$ depends only on  $h_{\a \beta}^+=\sum_{\ell,m} h_{\a \beta}^{(\ell,m,+)}$
\item There is a one to one relation between the gauge class $[h_{\a \beta}]$ of a metric perturbation 
(that is, the set of perturbations obtained from $h_{\a \beta}$ by a transformation (\ref{gt2})) and the 
 $ G^\pm$. More precisely, the maps
  \begin{equation}
[h_{\a \beta}^-] \to G_-([h_{\a \beta}^-] ), \;\;\; [h_{\a \beta}^+] \to G_-([h_{\a \beta}^+] )
\end{equation}
are bijections. In particular, the gauge perturbation in, say, the RW gauge, can be recovered from 
the $G_\pm$ fields. 

\end{enumerate}

For the odd sector of the LEE, a four dimensional approach relating the metric perturbation  with 
a scalar potential  $\Phi$ defined on the spacetime $\mathcal{M}$, instead of the $(t,r)$ orbit manifold, was 
found in \cite{Dotti:2013uxa}. It was noticed that the sum over $(\ell,m)$ of (\ref{rws}) simplifies to 

\begin{equation} \label{CsP}
{}^{RW}h_{\a \beta}^-= \sum_{(\ell \geq 2,m)} \mathcal{D}_{\a \beta}^{(\ell,m,-)}
\left[\phi_{(\ell,m)}^{-},S_{(\ell,m)}\right] = \frac{r^2}{3M} {}^* C_{\a}{}^{\g \delta}{}_{\beta} \nabla_{\g} \nabla_{\delta}
 \left( r^3 \Phi \right),
\end{equation}
where ${}^* C_{\a \beta \gamma \delta}= \tfrac{1}{2} \varepsilon_{\mu \nu \gamma \delta} C^{\alpha \beta}{}^{\mu \nu}$ 
is the dual of the Weyl tensor, and $\Phi: \mathcal{M} \to \mathbb{R}$ is a field assembled using spherical harmonics and the $\phi^{(\ell,m,-)}$:

\begin{equation} \label{4DRf}
 \Phi = \sum_{(\ell \geq 2,m)} \frac{\phi^{(\ell,m,-)}}{r} S_{(\ell,m)} : \mathcal{M} \to \mathbb{R}. 
\end{equation}

The odd sector LEE equations   for $\phi^{(\ell,m,-)}$, 
combined with  the spherical harmonic Equation (\ref{sh})  for the  $S_{(\ell,m)}: S^2 \to \mathbb{R}$ 
(see (\ref{1+1}) and (\ref{poten1})),  turn out 
to be equivalent to what we call the \textit{four dimensional Regge-Wheeler equation} (4DRWE),  which reads 

\begin{equation}  \label{4DRWE}
\nabla^{\a} \nabla_{\a} \Phi + \left( \frac{8M}{r^3} - \frac{2 \Lambda}{3}  \right) \Phi =0. 
\end{equation}

Note, however, that  $\Phi$ is no more than the collection of fields $\phi^-_{(\ell,m)}$: its connection to geometrically relevant 
fields such as CSs is, a priori,  loose.   Note also that \textit{any field} $\Psi: M \to \mathbb{R}$ of the form 
\begin{equation} \label{otf}
\Psi =  \sum_{(\ell \geq 2,m)} c_{(\ell,m)} \frac{\phi^{(\ell,m,-)}}{r} S^{(\ell,m)} 
\end{equation}
satisfies, for arbitrary constants $c_{(\ell,m)}$,  the 4DRWE (\ref{4DRWE}). 
This is a consequence of the form of the potential $V^{(\ell,m,-)}$ (see (\ref{poten1})), 
which contains an  ``angular momentum'' term $\ell(\ell+1)/r^2$ (contrast with $V^{(\ell,m,+)}$)

Much more important is the fact (proved in \cite{Dotti:2013uxa} for $\Lambda=0$, generalized to nonzero $\Lambda$ in 
\cite{dottiCQG}) 
that 
 the LEE implies that   the field $r^5 \dot Q_- = 
r^5 G_-$  can be written, after reiterated use of the LEE and equations derived from these, as
\begin{equation} \label{qdt}
G_- = \delta Q_- = - \frac{6M}{r^7} \sqrt{\tfrac{4 \pi}{3}}  \sum_{m=1}^3 j^{(m)}S^{(\ell=1,m)} - \frac{3M}{r^5} \sum_{\ell \geq 2, m} \frac{(\ell+2)!}{(\ell-2)!} 
\frac{\phi^{(\ell,m,-)}}{r} S^{(\ell,m)}
\end{equation}

The first term contains the $\ell=1$ contribution which, as we commented above, is time independent and  irrelevant to the 
stability problem (there is no $\ell=0$ odd contribution).
 This time independent contribution amounts to  a shift of the Schwarzschild BH to a ``slowly rotating Kerr black hole''. Technically, 
by ``slowly rotating Kerr black hole'' we mean the metric we get if we  Taylor expand Kerr's metric (\ref{kerr}) and (\ref{kerr2}) 
 in the rotation parameter 
$a$ and keep only first order terms: for such a perturbation, $G_-$ looks exactly like the $m=3$ (rotation around the $z-$ axis) 
term in the first sum in (\ref{qdt}). The second term of $r^5 G_-$, being of the form (\ref{otf}), satisfies the 4DRWE (\ref{4DRWE}). 
What is not obvious, but can be 
checked by direct calculation, is that the $\ell=1$ piece of $r^5 G_-$  (and thus $r^5 G_-$)  \textit{also} 
satisfies  (\ref{4DRWE}). $G_-$ in~(\ref{qdt}) contains all the 
information we need to reconstruct the metric perturbation: the $\phi^{(\ell,m,-)}$ can be obtained as the $\ell \geq 2$ harmonic 
coefficients of $r^5 G_-$, and the $j_m$ (which are all we need to reconstruct the $\ell=1$ piece of the metric perturbation, see \cite{dottiCQG}) 
can be obtained from  the $\ell=1$ coefficients. This proves that the gauge invariant, curvature scalar $G_-$ \textit{contains 
all the information  encoded in an odd metric perturbation, while being a meaningful scalar to measure the strength of the perturbation.} 
If we managed to place a pointwise bound on this quantity we would have a proof of nonmodal linear stability of the Schwarzschild BH 
under odd perturbations. 
A pointwise bound can be placed  on $G_-$  using  the fact that $r^5 G_-$  satisfies (\ref{4DRWE}),  by adapting a technique 
from \cite{Kay:1987ax} (see \cite{Dotti:2013uxa, dottiCQG} for details). 
The result is the following: for all $r>2M$ in the $\Lambda=0$ BH (or $r$ between the event and cosmological 
horizons if $\Lambda>0$), and all $t$, there is a constant $K_-$ that depends on the initial datum such~that
\begin{equation}\label{gmb}
G_- \leq \frac{K_-}{r^6}.
\end{equation}

This equation settles the issue of nonmodal stability under odd perturbations.

To treat even perturbations, we must face the problem that, as becomes obvious when inspecting the even potential
$V^{(\ell,m,+)}$ in (\ref{poten1}), particularly the factors involving $\ell$ in the denominator, the even Equation (\ref{1+1}) 
cannot be used together with (\ref{sh}) to construct a scalar field $\mathcal{M} \to \mathbb{R}$ that satisfies  a covariant equation 
such as the (\ref{4DRWE}). 
A remarkable  result by Chandrasekhar \cite{ch1, ch2}  comes to our help: there exists operators $\mathcal{D}_\ell^{\pm}$ of the form 

\begin{equation}\label{dpm}
\mathcal{D}_\ell^{\pm} = \pm \p_x + W_\ell = \pm f \p_r + W_\ell
\end{equation}
where
\begin{equation}\label{dpm1}
W_{\ell} = \frac{1}{12M} \frac{(\ell+2)!}{(\ell-2)!} + \frac{6M f(r)}{r \; (\mu r + 6M)},
\end{equation}
such that, if $\phi^{(\ell,m,-)}$ is a solution of the odd Equation (\ref{1+1}), then $\mathcal{D}_\ell^{+} \phi^{(\ell,m,-)}$ is 
a solution of the even  equation, and similarly replacing $+ \leftrightarrow -$. In \cite{dottiCQG}, Lemma 7, it was furthermore proved that, although 
the operators (\ref{dpm}) clearly have a non trivial kernel \textit{when acting on arbitrary functions}, they are 1-1 when restricted 
to \textit{solutions of the 1+1 wave Equations (\ref{1+1})}. As a consequence, \text{every} solution $\phi^{(\ell,m,+)}$ 
of the even 1+1 wave equation in the 
$r>2M$ domain 
of a $\Lambda = 0$   Schwarzschild BH ($r$ between the event and cosmological horizons if $\Lambda>0$), 
can be written as $\phi^{(\ell,m,+)}= \mathcal{D}_\ell^+ \phi^{(\ell,m,-)}$.

Using the even LEE and equations derived from those, $G_+$  can be reduced to \cite{Dotti:2013uxa} 

\begin{equation} \label{G+}
G_+ = - \frac{2M\;  \delta M}{r^5} + \frac{M}{2r^4} \sum_{\ell \geq 2} \frac{(\ell+2)!}{(\ell-2)!} \left[ f \p_r + Z_{\ell} \right] \phi^{(\ell,m,+)} S_{(\ell,m)},
\end{equation}
where 
\begin{equation} 
Z_{\ell}  = \frac{2M \Lambda r^3+\mu r (r-3M)-6M^2}{r^2 (\mu r + 6M)}, \;\; \mu = (\ell-1)(\ell+2).
\end{equation}

The---time-independent---first term in (\ref{G+}) comes from  the $\ell=0$ even perturbation, which amounts to 
a  mass shift of the background metric \cite{dottiCQG} (there is no $\ell=1$ even contribution, see \cite{dottiCQG}). 
Further use of the Chandrasekhar operators described above  allows to rewrite $G_+$ entirely in terms of functions obeying  (\ref{4DRWE}). 
The details, which are quite involved, as well as the details of how to use this fact to set a bound on $G_+$, can be found in \cite{dottiCQG}.
We only quote the  result obtained in \cite{Dotti:2013uxa,dottiCQG}: 
for all $r>2M$ in the $\Lambda=0$ BH ($r$ between the event and cosmological 
horizons if $\Lambda>0$) and all $t$, there is a constant $K_+$ that depends on the initial datum such that 
\begin{equation}
G_+ \leq \frac{K_+}{r^4}.
\end{equation}

Together with (\ref{gmb}), this equation proves the nonmodal linear stability of $\Lambda \geq 0$ Schwarzschid BHs.

We close this sections with two observations made in \cite{dottiCQG}. 
Combining (\ref{qdt}) and (\ref{G+}) with the Price tail decay at fixed $r$, Equation (\ref{pt}), we find that, 
at large $t$ and fixed $(r,\theta,\phi)$,

\begin{equation} \label{Gap}
G_- \simeq - \frac{6M}{r^7} \sqrt{\tfrac{4 \pi}{3}}  \sum_{m=1}^3 j^{(m)}S^{(\ell=1,m)}, \;\;\; 
G_+ \simeq - \frac{2M\;  \delta M}{r^5},
\end{equation}
which corresponds to a stationary BH in the Kerr (or Kerr de Sitter) family with a mass $M + \delta M$ and 
angular momentum components $j^{(m)}$. For example,  if only $j^{(3)}\neq 0$, the perturbation (\ref{Gap}) 
corresponds (in some gauge) to the metric perturbation obtained  by applying the operator $\delta M \; \p_M + j^{(3)}/M \; \p_a$ 
to the metric  (\ref{kerr}) and (\ref{kerr2}). 
Equation (\ref{Gap}) indicates  that, after a long time, the perturbation settles into a ``slowly rotating'' Kerr (or Kerr dS) BH. 
For $\Lambda=0$, this fact was rigorously proved in  \cite{Daf1}, where it was shown, working in an specific gauge, that 
  the metric perturbation decays at large $t$ into that of a slowly rotating~BH.

  The second observation is that there is a 
  much simpler CS connected to even perturbations, but this CS is not gauge invariant. This is $Q_+$, defined in (\ref{curvaturescalars}) which, 
  \textit{in the Regge-Wheeler gauge}, has a first order variation \cite{dottiCQG}
    \begin{equation} \label{qp0}
\delta Q_+^{(RW)} = \frac{2M \; \delta M}{r^6} - \frac{6M^2}{r^5} \sum_{(\ell\geq 2,m)} \frac{\mathcal{D}^-_{\ell} \phi^{(\ell,m,+)}}{r} S^{(\ell,m)}.
\end{equation}

Thus, $r^5  \delta Q_{+,>1}^{(RW)}$, being of the form (\ref{otf}),  also satisfies the 4DRW equation. We might think of using  $\delta Q_+$ 
to measure the strength of even perturbations, but 
this field, as we said,  
 is gauge dependent,  due to the fact that  $Q_+ \neq 0$ for the background Schwarzschild or S(A)dS black hole and, as a consequence, 
under the gauge transformation (\ref{gt2}),
\begin{equation} \label{qp1}
\delta Q_+ \to {\delta Q_+}{}' = \delta Q_+ +  \zeta^r \p_r Q_+ = \delta Q_+ -  \zeta^r  \; \frac{6M^2}{r^7}.
 \end{equation}
 
 A gauge invariant field $H_+: \mathcal{M} \to \mathbb{R}$
  was found in \cite{dottiCQG} for the even perturbations which satisfies the 4DRWE  (\ref{4DRWE}) (see
 the discussion in Section 5.2 of this reference). 
 This has the drawback of not having a direct geometric interpretation in terms of CSs. In any case, we learn 
 from the existence of $H_+$, or just from Equation (\ref{qp0}), that the degrees of freedom of the 
 linearized even perturbations can  be encoded in a scalar field $\mathcal{M} \to \mathbb{R}$ satisfying the 4DRWE 
 (\ref{4DRWE}), as is the case for the odd perturbations.   This field  is independent from the odd field (\ref{4DRf}), that also 
 satisfies this equation. In other words: \textit{the most general linear perturbation can be 
 encoded in two independent scalar fields $\mathcal{M} \to \mathbb{R}$ which satisfy (\ref{4DRWE})}. 
 Thus, proving the stability and the large $t$ decay of perturbations of a Schwarzschild or Schwarzschild-de Sitter 
 BH into slowly rotating Kerr (de Sitter) BH amounts to proving the decay of the $\ell \geq 2$ components 
of generic solutions of  (\ref{4DRWE}). This is at the heart of the proof in \cite{Daf1} of the decay of perturbations for $\Lambda=0$, where the two scalars 
satisfying (\ref{4DRWE}) were integrated into an $S^2$ symmetric tensor, as explained in Remark 7.1 in \cite{Daf1}.

\section{Conclusions and Current Developments}

The proof of the instabilities of the naked singularities and  the regions beyond the black hole 
 Cauchy horizon within the Kerr--Newman family of metrics, conciliates 
General Relativity with basic Physics principles, such as the
 uniqueness of evolution from initial data, and the requirement that  causal pathologies such as closed timelike and null  curves 
 do not occur.  
A modal analysis of the LEE is enough to rule out these solutions, 
given that exponentially growing modes are found. It is an interesting fact that, in the non rotating case, 
 the geometrical effects of the unstable 
modes show up in  \textit{differential} curvature scalar invariants 
and in the splitting of one of the background degenerate Petrov null directions (see Equations (\ref{este}) and (\ref{ls})). They do not 
leave traces on \textit{algebraic curvature scalars}, in contrast to what happens   for generic black hole perturbations, (see, for example, 
Equations~(\ref{qdt}) and (\ref{G+})). 
 It is also interesting that the unstable modes in all the analyzed static cases (negative 
mass Schwarzschild NS, super extreme Reissner-Nordström NS, inner region of the Reissner-Nordström BH) are  even under $P$, as defined in 
(\ref{ps}), and that there is exactly one unstable mode in each harmonic sector. 

An analysis 
of these cases suggests that the singularity plays no role in the instability. As an example, for the Schwarzschild naked singularity, a
 negative sector of the potential $V^{(\ell,m,+)}$ in the 
1+1 wave Equations (\ref{1+1})--(\ref{poten1}) is responsible for the existence of the unstable modes. If the singularity were replaced by a small 
 spherically symmetric 
negative mass distribution, away from $r=0$ this potential would not change substantially, and an unstable mode could be found anyway. 
This is also the case of the super-extreme Reisner-Nordström NS, treated in \cite{doglepu, Dotti:2010uc}, for which the singularity could be replaced by 
a spherically symmetric charge distribution with $|Q|>M$ leading to a non singular metric which would, anyway, be unstable. 
Similarly,   for the Kerr NS, the ring singularity plays no role in its instability. 
 This leaves the impression that General Relativity simply ``dislikes'' unusual matter or 
over-rotating and over-charged compact objects.

The analysis of the stability of  black hole  outer regions, having passed decades ago the  test of modal linear stability,  
has had,  after a long period of little activity, a remarkable 
 progress in the last few years. An incomplete list of recent advances related to  the stability of the  Schwarschild and Kerr black holes 
 follow: (i) for $\Lambda \geq 0$ Schwarzschild black holes the nonmodal linear stability was established in \cite{Dotti:2013uxa,dottiCQG};
(ii) in the $\Lambda=0$ case, the \textit{decay} in time 
of generic linear perturbations of the Schwarzschild black hole, leaving  a  ``slowly'' rotating Kerr black hole  was proved 
in \cite{Daf1}; (iii) the conditional 
stability of the $\Lambda<0$ Schwarzschild black hole, and the breaking of the even/odd symmetry mediated by 
the Chandrasekhar operators (\ref{dpm}) and (\ref{dpm1}) was studied in \cite{Araneda2015}; 
{(iv)} the \textit{non-linear} stability of the Schwarzschild de Sitter black hole was  proved in~~\cite{hv};  {(v)} % 2 iv, please revise FIXED
a preprint is now available with a proof of the \textit{non-linear} stability of the $\Lambda=0$ 
Schwarzschild black hole \cite{Daf2};
(vi) pointwise decay estimates for solutions of the linearized Einstein's equations  on the  outer region  of a Kerr black hole were obtained in 
\cite{lars}; (vii) the role of hidden symmetries (see the review~\cite{Frolov2017}) 
in type D  spacetimes, and the reconstruction of (gravitational, Maxwell and spinor) perturbation fields 
from ``Debye potentials'' (first introduced in \cite{Kegeles,wald2}),  was studied in depth  and made clear in the series of papers 
\cite{Araneda2016,Araneda2017,Araneda2018,Araneda2020}.

The ultimate challenge in Black Hole perturbation theory remains open: proving the non-linear stability of the outer region
 of a Kerr black hole.

\acknowledgments{This review is an outgrowth of the lecture notes for a course on linear stability 
of black holes and naked singularities delivered at the V Jos\'e Pl\'{\i}nio Baptista School of Cosmology, 
held at Guarapari (Esp\'{\i}rito Santo) Brazil, from 30 September to 5 October 2021. I thank the organizers for giving me the opportunity of lecturing  at the School. This research was funded by CONICET (Argentina) grant PIP 11220080102479 and Universidad Nacional de Córdoba (Argentina), 
grant 30720110 101569CB.}

%Declare conflicts of interest or state ``The authors declare no conflict of interest.'' Authors must identify and declare any personal circumstances or interest that may be perceived as inappropriately influencing the representation or interpretation of reported research results. Any role of the funders in the design of the study; in the collection, analyses or interpretation of data; in the writing of the manuscript, or in the decision to publish the results must be declared in this section. If there is no role, please state ``The funders had no role in the design of the study; in the collection, analyses, or interpretation of data; in the writing of the manuscript, or in the decision to publish the~results''.

% Please provide either the correct journal abbreviation (e.g., according to the “List of Title Word Abbreviations” http://www.issn.org/services/online-services/access-to-the-ltwa/) or the full name of the journal.

%\externalbibliography{yes}
%\bibliography{your_external_BibTeX_file}

%=====================================
% References, variant B: internal bibliography
%=====================================

\end{document}